\documentclass[14pt,reprint,twocolumn,longbibliography,
superscriptaddress,
amsmath,
amssymb,
pra
]{revtex4-2}
\usepackage{graphicx,amsmath,subfigure}%
\usepackage{color,xcolor}
\usepackage[bookmarks=false]{hyperref}
\hypersetup{colorlinks=true,citecolor=cyan,linkcolor=blue,urlcolor=blue,pdfstartview=FitH,bookmarksopen=true}
\usepackage{epstopdf}
\usepackage{float}
\usepackage{ulem}
\usepackage{upgreek}
\usepackage{silence}
\usepackage{inputenc}\inputencoding{utf8}\DeclareUnicodeCharacter{200B}{}
\begin{document}

\title{Time-Frequency Correlated Native CCZ Gate in Superconducting Circuits}

\author{Chenhui Wang}
\email{ccchui123@outlook.com}
\affiliation{Laboratory for Advanced Computing and Intelligence Engineering, Zhengzhou 450001, Henan, China}
\author{Weilong Wang} 
\affiliation{Laboratory for Advanced Computing and Intelligence Engineering, Zhengzhou 450001, Henan, China}
\author{Yangyang Fei} 
\affiliation{Laboratory for Advanced Computing and Intelligence Engineering, Zhengzhou 450001, Henan, China}
\author{Zhiqiang Fan} 
\affiliation{Laboratory for Advanced Computing and Intelligence Engineering, Zhengzhou 450001, Henan, China}
\author{Hanshi Zhao} 
\affiliation{Laboratory for Advanced Computing and Intelligence Engineering, Zhengzhou 450001, Henan, China}
\author{Geyuyan Ma} 
\affiliation{Laboratory for Advanced Computing and Intelligence Engineering, Zhengzhou 450001, Henan, China}
\author{Zheng Shan}
\email{shanzhengzz@163.com}
\affiliation{Laboratory for Advanced Computing and Intelligence Engineering, Zhengzhou 450001, Henan, China}
\begin{abstract}
	Practical quantum advantage hinges on executing deep quantum circuits within the coherence limits of noisy intermediate-scale quantum processors. The absence of native, high-fidelity multi-qubit gates remains a major bottleneck, as their decomposition into single- and two-qubit gates leads to prohibitive depth and error overhead. Here, we propose a hardware-efficient protocol that directly implements a native controlled-controlled-Z (CCZ) gate in a tunable-coupler superconducting circuit. Our theoretical protocol activates a resonant three-qubit interaction via a time-frequency correlated virtual process, explicitly relying on the dynamic resonant exchange within the $|101\rangle \leftrightarrow |020\rangle$ transition manifold. This approach is compatible with standard tunable-coupler architectures without requiring additional control resources. Through a systematic calibration workflow combining pulse shaping and active cancellation of residual phases, we demonstrate a gate fidelity exceeding 99\% within $165\,\mathrm{ns}$---significantly outperforming decomposed sequences. Comprehensive error budgeting confirms that the gate performance remains robust against  realistic experimental imperfections. Furthermore, we show that this scheme can be naturally extended to a continuous $\mathrm{CCPhase}(\theta)$ gate set. This work provides a direct, high-fidelity route to three-qubit entanglement, offering promising prospects for efficient execution of quantum algorithms on near-term superconducting hardware.
\end{abstract}
\maketitle
\section{Introduction}
While superconducting quantum computing has witnessed remarkable progress in scale and performance over the past decade~\cite{Arute2019,Acharya2025,rqkg-dw31}, practical quantum advantage remains stifled by the prohibitive depth of compiled circuits relative to available coherence times~\cite{Preskill2018quantumcomputingin,Yamasaki2024}. 
Although single-qubit and two-qubit gates constitute a universal set~\cite{PhysRevA.52.3457}, synthesizing multi-qubit logic from these primitives incurs substantial overhead~\cite{PhysRevA.106.042602,PhysRevA.109.052440}. 
The three-qubit controlled-controlled-Z (CCZ) gate is particularly crucial for Grover's search~\cite{PhysRevLett.79.325,lvb9-pfr3}, Shor's algorithm~\cite{Gidney2021howtofactorbit}, and quantum error correction~\cite{tasler2025optimizingsuperconductingthreequbitgates}. 
Standard decompositions typically necessitate a cascade of eight CNOT gates interspersed with single-qubit rotations~\cite{lvb9-pfr3}. 
This cascade compounds error rates and prolongs exposure to decoherence. 
Consequently, a native high-fidelity CCZ gate is essential for compressing circuit depth in the noisy intermediate-scale quantum (NISQ) era.

This imperative has driven extensive research into direct three-qubit gate synthesis across diverse platforms: Rydberg atom arrays leverage blockade engineering~\cite{k72m-9tn8}, trapped ions utilize auxiliary qutrit levels~\cite{Figgatt2019}, and silicon employs anisotropic exchange interactions in triple-dot arrays~\cite{kp8s-py9m}. In superconducting circuits, high-fidelity three-qubit gates often demand specialized hardware designs. Such as fluxonium-transmon hybrid architectures that mitigate cross-talk and frequency crowding~\cite{PhysRevApplied.21.044035}---which introduce considerable fabrication complexity and scalability challenges. Recent tunable-coupler implementations~\cite{lvb9-pfr3} realize the CCZ gate primarily by exploiting interactions within the three-excitation manifold (e.g., involving the $|111\rangle$ state) or relying on adiabatic flux-biasing trajectories. However, accessing these high-excitation or adiabatic pathways inherently suffers from limited tunable coupling ranges~\cite{Lange_2026} and severe residual $ZZ$-crosstalk~\cite{PhysRevX.11.021058,fors2024comprehensiveexplanationzzcoupling} that necessitates intricate calibrations. Crucially, the strict requirement for adiabaticity and the heightened decoherence vulnerability of higher energy levels fundamentally constrain the gate speed (typically requiring $\sim$300~ns), thereby limiting the executable circuit depth within the finite coherence time. Consequently, a hardware-efficient, high-fidelity CCZ gate compatible with mainstream superconducting architectures remains a critical unmet need.

To circumvent these limitations, we propose a hardware-efficient scheme that strictly confines the gate dynamics to the two-excitation manifold. In striking contrast to adiabatic or $|111\rangle$-dependent protocols, we realize a native CCZ gate by engineering a fast, time-frequency correlated virtual transition directly between the $|101\rangle$ and $|020\rangle$ states. Analogous to optical two-photon absorption~\cite{goppert1931elementarakte}, the central transmon's $|2\rangle$ state mediates a resonant three-body coupling~\cite{PhysRevLett.125.133601,PhysRevLett.128.190502,Aamir2025}. Benefiting from this fast resonant exchange, our QuTiP-based open-system simulations~\cite{johansson2012qutip} demonstrate a gate fidelity exceeding $99\%$ within just $165\,\mathrm{ns}$. Validated against realistic noise and extended to continuous CCPhase($\theta$) gates, this approach provides a robust and practical route to dense quantum logic.

The paper is organized as follows: In Sec.~\ref{Sec2}, we establish the theoretical framework for the tunable superconducting circuit, derive the effective three-body Hamiltonian via the Schrieffer-Wolff transformation, and elucidate the resonant physical mechanism. Section~\ref{Sec3} details the implementation strategy, presenting a calibration protocol that covers the geometric phase accumulation, the active cancellation of residual dynamical errors, and the extension to continuous CCPhase($\theta$) gates. In Sec.~\ref{Sec4}, we characterize the gate performance through analysis of the realized unitary evolution, a mechanistic error budget breakdown, and a systematic robustness analysis. Finally, we conclude the work and discuss future outlooks in Sec.~\ref{Sec5}.

\section{physical model of the system}\label{Sec2}

\subsection{Circuit Hamiltonian and System Setup}

\begin{figure}[t]
	\centering
	\includegraphics[width=\linewidth]{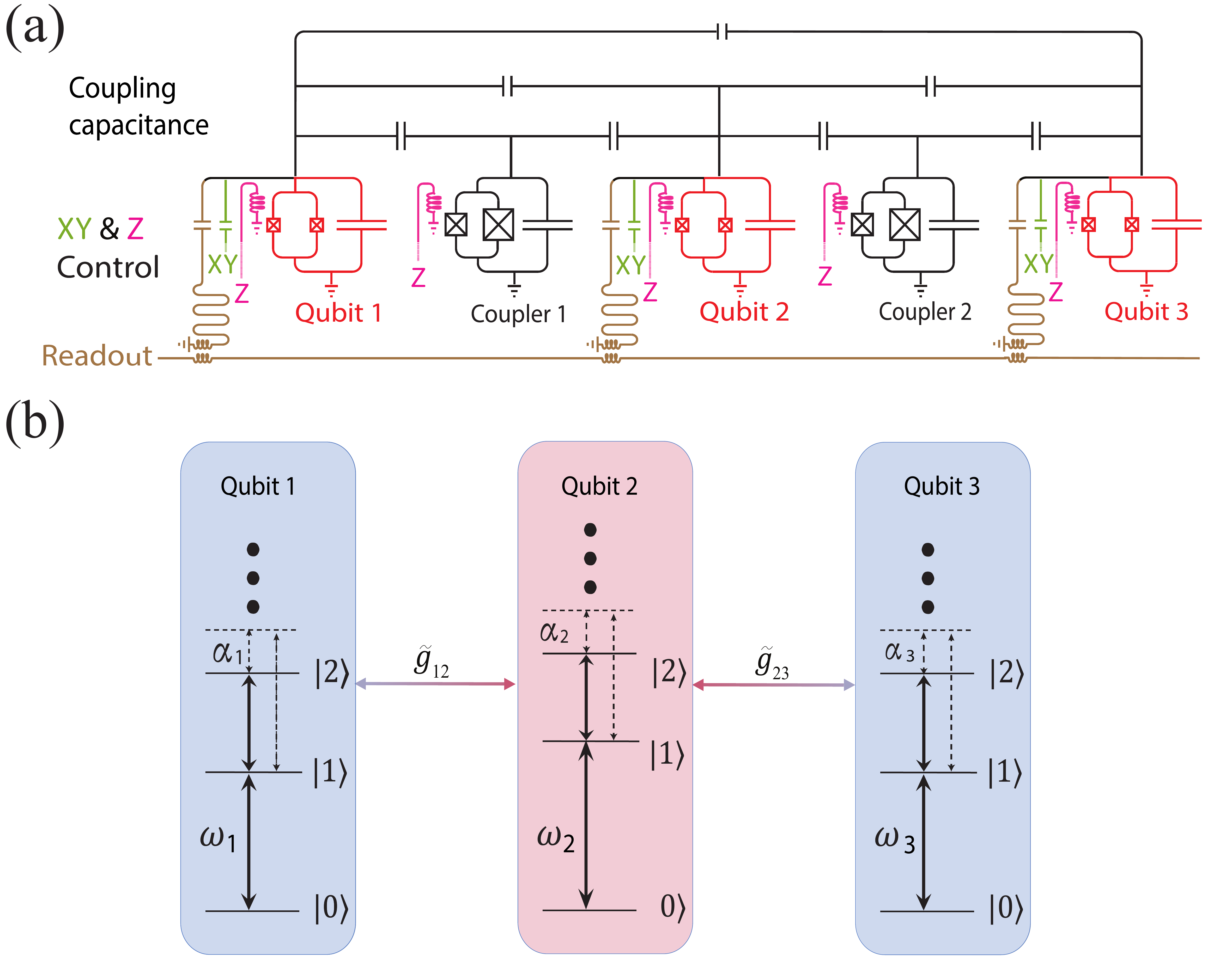}
	\caption{
		(a) Circuit schematic of the superconducting processor. Three transmon qubits (denoted as $Q_1$--$Q_3$) are coupled via tunable couplers (denoted as $C_1$, $C_2$). Each qubit has independent XY and Z control lines, alongside a dedicated readout resonator.
		(b) Effective energy-level diagram obtained after a Schrieffer-Wolff transformation that adiabatically eliminates the couplers, yielding effective nearest-neighbor couplings $\tilde{g}_{12}$ and $\tilde{g}_{23}$. The second excited state $|2\rangle$ of $Q_2$ serves as the auxiliary mediator for the three-body interaction. The anharmonicity $\alpha_i$ (energy difference between the dashed and solid lines extending from $|1\rangle$) sets the precise energy of the $|2\rangle$ state. A direct capacitive coupling $g_{13}$ between $Q_1$ and $Q_3$ is designed to be negligible and is omitted from the effective model.
	}
	\label{fig:scheme}
\end{figure}

We consider a linear chain of three superconducting transmon qubits ($Q_1, Q_2, Q_3$) coupled via tunable couplers ($C_1, C_2$) as illustrated in Fig.~\ref{fig:scheme}(a). Both qubits and couplers are flux-tunable transmons, which are essentially nonlinear superconducting LC circuits characterized by their Josephson energy $E_J$ and charging energy $E_C$. Under the condition $E_J \gg E_C$, these devices are well described as weakly anharmonic Duffing oscillators~\cite{PhysRevA.76.042319}. The five-qubit system can be described by the following Hamiltonian (setting $\hbar = 1$ throughout this paper):
\begin{equation}
	\label{eq:hami}
	\begin{aligned}
		\hat{H}_{\mathrm{sys}} = \sum_{k \in \{1,2,3,c_1,c_2\}} \left( \omega_k \hat{a}^\dagger_{k} \hat{a}_{k} + \frac{\alpha_k}{2} \hat{a}^\dagger_{k} \hat{a}^\dagger_{k} \hat{a}_{k} \hat{a}_{k} \right) \\
		+\sum_{\langle i,j \rangle} g_{ij} (\hat{a}^\dagger_i - \hat{a}_i)(\hat{a}_j - \hat{a}^\dagger_j),
	\end{aligned}
\end{equation}
where $\hat{a}_{k}$ ($\hat{a}^\dagger_{k}$) is the annihilation (creation) operator for mode $k$ with transition frequency $\omega_k$ and anharmonicity $\alpha_k$. The coefficient $g_{ij}$ are proportional to the bare capacitive coupling strength between adjacent qubits. While tuning the Josephson energy $E_J$ of $Q_2$ physically introduces a weak time dependence to the bare couplings ($g_{ij} \propto \sqrt{\omega_i \omega_j}$), the fractional variation is negligible ($\sim 1\%$) given the limited tuning range. The system dynamics are dominated by the time-dependent detunings in the effective couplings. Thus, following standard practice, we treat the bare $g_{ij}$ as static parameters in our model. The frequencies $\omega_k$ can be tuned dynamically by an external magnetic flux threading the SQUID loops~\cite{PhysRevApplied.8.044003}.

To derive a tractable effective model, we work in the dispersive regime where the qubit-coupler detuning significantly exceeds the coupling strength ($|g_{i,c_j}| \ll |\omega_i - \omega_{c_j}|$). In this limit, the couplers serve as virtual mediators while remaining predominantly in their ground states. We apply a Schrieffer-Wolff transformation (SWT)~\cite{BRAVYI20112793} to adiabatically eliminate the coupler degrees of freedom and obtain an effective Hamiltonian for the three computational qubits:
\begin{equation}
	\label{eq:H2}
	\begin{aligned}
		\hat{H}_{\mathrm{eff}} = \sum_{i=1}^{3} \Bigl( \tilde{\omega}_i \hat{a}^\dagger_i \hat{a}_i + \frac{\alpha_i}{2} \hat{a}_i^\dagger \hat{a}_i^\dagger \hat{a}_i \hat{a}_i \Bigr) \\
		+ \tilde{g}_{12}(t) (\hat{a}^\dagger_1 \hat{a}_2 + \hat{a}_1 \hat{a}^\dagger_2) + \tilde{g}_{23}(t) (\hat{a}^\dagger_2 \hat{a}_3 + \hat{a}_2 \hat{a}^\dagger_3),
	\end{aligned}
\end{equation}
where $\tilde{\omega}_i$ is the Lamb-shifted qubit frequency, and $\tilde{g}_{ij}(t)$ is the effective nearest-neighbor coupling strength mediated by the couplers. Detailed derivations of these parameters via SWT are given in Appendix~A. The coupling $\tilde{g}_{ij}$ is set by the detuning between the couplers and the qubits. By tuning the coupler frequency, $\tilde{g}_{ij}$ can be varied continuously from positive to negative values, crossing zero as demonstrated in Ref.~\cite{PhysRevApplied.10.054062}. Direct next-nearest-neighbor coupling $g_{13}$ is structurally suppressed ($|g_{13}| \ll |\tilde{g}_{ij}|$) to minimize crosstalk and is therefore omitted.

In our numerical simulations, we evolve the system under the time-dependent Hamiltonian in Eq.~\eqref{eq:H2}, modulating $\tilde{\omega}_i(t)$ and $\tilde{g}_{ij}(t)$ via the pulse sequences described in Sec.~\ref{Sec3}. To capture dominant leakage while maintaining computational efficiency, we truncate the local Hilbert space of each transmon to its three lowest energy levels (the qutrit approximation). The collective states are denoted by $|q_1 q_2 q_3\rangle$ with $q_i \in \{0,1,2\}$. This truncation is justified by the spectral selectivity of our shaped pulses and the fact that the transmon anharmonicity is sufficiently large compared to the inter-qubit coupling strengths ($|\alpha_i| \gg |g_{ij}|$). Its validity is confirmed by simulating the system with an extended truncation of 10~levels per transmon and quantifying the residual leakage, as detailed in the error analysis of Sec.~\ref{Sec4}.

\begin{figure}[t]
	\centering
	\includegraphics[width=\linewidth]{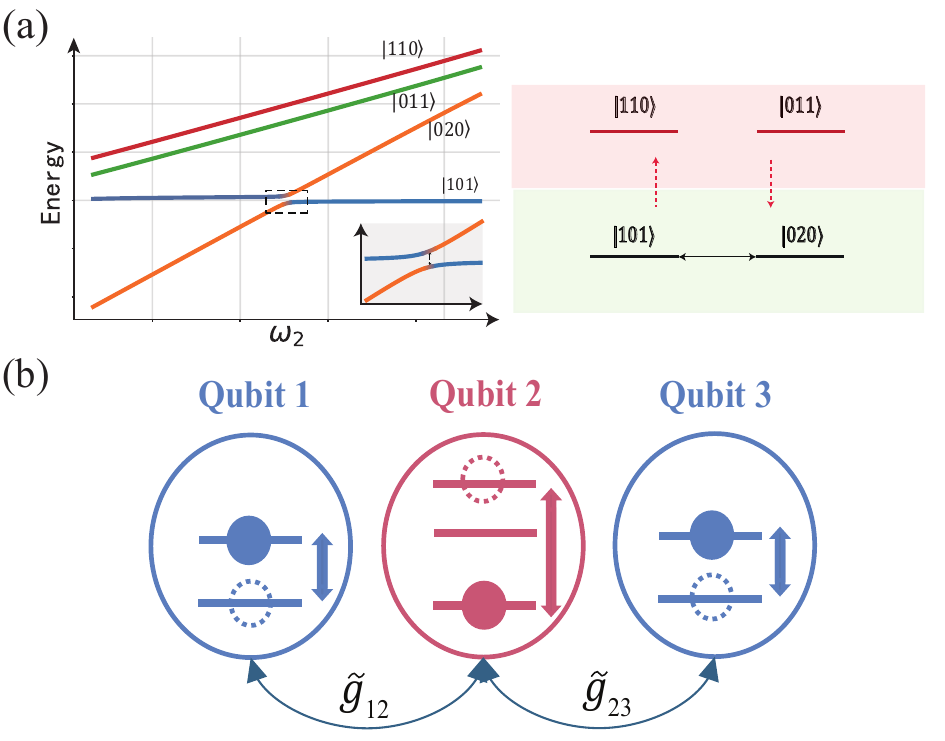}
	\caption{\label{fig:mechanism}
		(a) Eigenenergy spectrum near the resonance condition $\omega_1 + \omega_3 \approx 2\omega_2 + \alpha_2$, showing the avoided crossing between the computational state $|101\rangle$ and the auxiliary state $|020\rangle$. The interaction is mediated virtually via the off-resonant intermediate states $|110\rangle$ and $|011\rangle$. The finite detuning results in a residual non-adiabatic population, a leading source of coherent error analyzed in Sec.~\ref{Sec4}.
		(b) Schematic of the resonant excitation-exchange process. The coupling drives a cyclic Rabi oscillation between $|101\rangle$ and $|020\rangle$, simultaneously annihilating single excitations in $Q_1$ and $Q_3$ while creating a double excitation in $Q_2$. This closed trajectory imparts a conditional geometric phase of $\pi$ to $|101\rangle$ upon its return to the computational subspace.
	}
\end{figure}

\subsection{Effective Interaction and Gate Dynamics}

To engineer the target three-body coupling, we utilize the second excited state $|2\rangle$ of the central qubit as an auxiliary mediator. In the idle configuration, the auxiliary state $|020\rangle$ is detuned from the computational state $|101\rangle$. We activate the interaction via a fast frequency shift of $Q_2$ to bring $\omega_2(t)$ into resonance:
\begin{equation}
	\label{eq:resonance}
	\omega_1 + \omega_3 \approx 2\omega_2(t) + \alpha_2 .
\end{equation}

This condition underpins our time-frequency correlated mechanism. As illustrated in the right panel of Fig.~\ref{fig:mechanism}(a), the direct transition between $|101\rangle$ and $|020\rangle$ is forbidden. Instead, the coupling is mediated through two distinct virtual pathways involving the intermediate single-excitation states $|110\rangle$ and $|011\rangle$. 

Physically, this process is analogous to a two-photon Raman transition. The simultaneous annihilation of single excitations in $Q_1$ and $Q_3$ is energetically compensated by the creation of a double excitation in $Q_2$. Because the intermediate states remain far detuned from the $|101\rangle \leftrightarrow |020\rangle$ resonance manifold (i.e., the detunings are much larger than the couplings $\tilde{g}_{ij}$), energy conservation prevents them from acquiring macroscopic real populations. Thus, they act solely as spectrally correlated virtual levels that temporally synchronize the excitation exchange.

To isolate the resonant dynamics, we partition the effective three-qubit Hamiltonian [Eq.~\eqref{eq:H2}] into an unperturbed diagonal part $\hat{H}_{\mathrm{diag}}(t)$ and an off-diagonal coupling part. The unperturbed Hamiltonian, which governs the free evolution and incorporates the time-dependent frequency tuning of $Q_2$, is explicitly given by
\begin{equation}
	\hat{H}_{\mathrm{diag}}(t) = \sum_{i=1}^{3} \left[ \tilde{\omega}_i(t) \hat{a}_i^\dagger \hat{a}_i + \frac{\alpha_i}{2} \hat{a}_i^\dagger \hat{a}_i^\dagger \hat{a}_i \hat{a}_i \right].
	\label{eq:H_diag_main}
\end{equation}

We then transform the system into the interaction picture defined by $U_I(t) = \exp\left[ i\int_0^t \hat{H}_{\mathrm{diag}}(\tau) d\tau \right]$ (which corresponds exactly to the rotating frame with respect to the unperturbed energies utilized in our Appendix). By adiabatically eliminating the transient amplitudes of the virtually populated states $|110\rangle$ and $|011\rangle$ (a rigorous derivation of the virtual dynamics is provided in Appendix~A), we obtain the effective dynamics for the target subspace $\{|101\rangle, |020\rangle\}$, as illustrated in the right panel of Fig.~\ref{fig:mechanism}(a). While this second-order elimination inherently generates state-dependent diagonal energy shifts (including single-qubit AC-Stark shifts and residual $ZZ$ cross-talks), the operational resonance is maintained by dynamically compensating the single-qubit detunings via our optimized control pulse in Stage 1. Furthermore, the accumulated multi-qubit dynamical phases are completely removed by the active cancellation protocol in Stage 2 (see Sec.~\ref{Sec3}B). Thus, the core coherent excitation exchange driving the geometric phase within the $\{|101\rangle, |020\rangle\}$ manifold (see the schematic illustration in Fig.~\ref{fig:mechanism}b) is faithfully captured by the effective pure-coupling Hamiltonian:
\begin{equation}
	\label{eq:Heff}
	\tilde{H}_{\mathrm{eff}} = J \hat{a}_1 \hat{a}_3 \hat{\Xi}^+ + \mathrm{H.c.},
\end{equation}
where $\hat{\Xi}^+ = |2\rangle\langle 0|$ is the transition operator to the second excited state of $Q_2$. The effective three-body coupling strength $J$ is
\begin{equation}
	\label{eq:J_eff}
	J \approx \sqrt{2}\,\tilde{g}_{12}\tilde{g}_{23}
	\left( \frac{1}{\tilde{\omega}_1 - \tilde{\omega}_2} + \frac{1}{\tilde{\omega}_3 - \tilde{\omega}_2} \right),
\end{equation}
with $\tilde{g}_{ij}$ and $\tilde{\omega}_i$ denoting the effective nearest-neighbor coupling and the dressed qubit frequency, respectively. This expression, derived in Appendix~A via the Schrieffer-Wolff transformation, underscores the hardware efficiency of our scheme: the three-body interaction $J$ is generated directly from native two-body couplings and controlled detunings, eliminating the need for any direct physical connection between $Q_1$ and $Q_3$.

The gate is realized by driving a complete Rabi oscillation along the trajectory $|101\rangle \to i|020\rangle \to -|101\rangle$ within the $\{|101\rangle, |020\rangle\}$ subspace. This imprints a conditional phase of $\pi$ exclusively on the $|101\rangle$ state upon its return to the computational subspace, thereby implementing a native controlled-controlled-phase operation. We note that this operation is locally equivalent to the standard CCZ gate—which applies a $\pi$ phase to $|111\rangle$—up to a single-qubit $X$ gate on $Q_2$. The geometric phase accumulated during this cyclic evolution (analyzed in the interaction picture above) becomes the observable conditional phase after transforming back to the Schrödinger picture and actively canceling deterministic dynamical phases, as detailed in Sec.~\ref{Sec3}.

\begin{figure}[t]
	\centering
	\includegraphics[width=\linewidth]{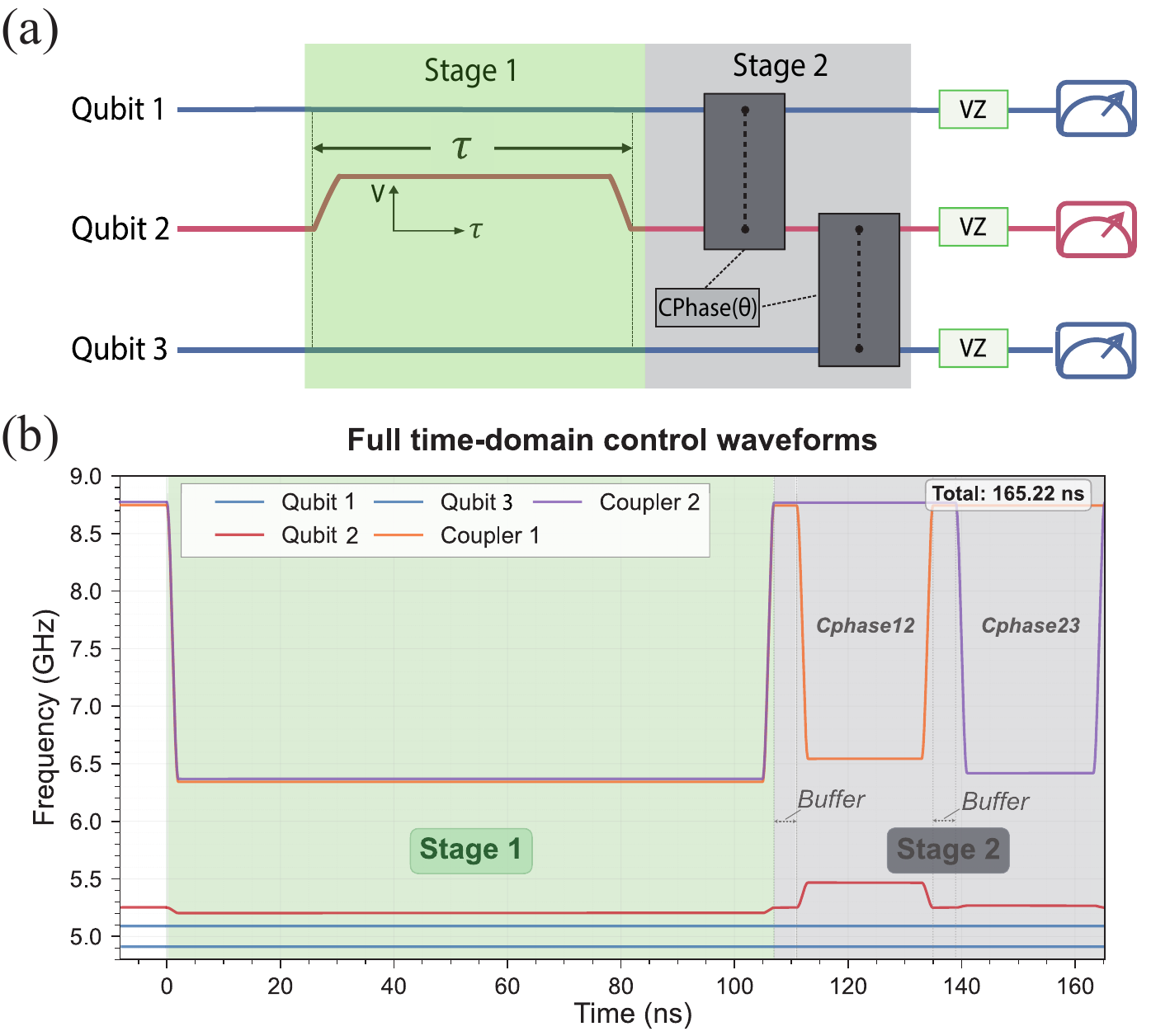}
	\caption{\label{fig:sequence}
		(a) Schematic of the two-stage gate sequence. In the first stage (geometric phase imprinting), a flat-top pulse of duration $\tau$ applied to $Q_2$ drives the system along the cyclic trajectory $|101\rangle \leftrightarrow |020\rangle$, accumulating the target geometric phase. In the second stage (active cancellation), short corrective CPhase gates (gray blocks) on the $Q_1$–$Q_2$ and $Q_2$–$Q_3$ pairs cancel residual dynamical phases from diagonal $ZZ$ interactions, while single-qubit phase shifts are corrected via software-implemented virtual-Z (VZ) frame updates.
		(b) Full time-domain control waveforms, showing the synchronized frequency pulse sequence applied to the qubits and couplers over the entire gate duration. The waveform includes the main resonant drive of Stage 1 followed by the compensation sequences of Stage 2. Fixed $5\,\mathrm{ns}$ idle buffers separate the stages to isolate them and suppress transient crosstalk.
	}
\end{figure}

\section{NUMERICAL VALIDATION OF THE IMPLEMENTATION STRATEGY}
\label{Sec3}

\begin{figure}[t]
	\centering
	\includegraphics[width=\linewidth]{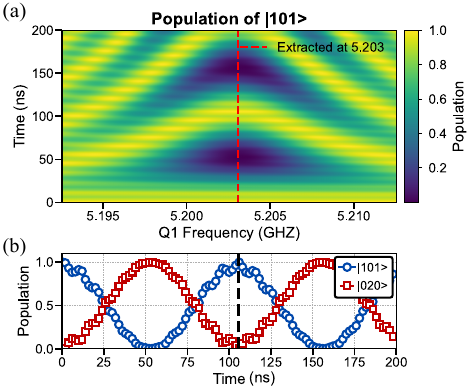}
	\caption{\label{fig:calibration}
		(a) Simulated population of $|101\rangle$ versus interaction duration and the frequency detuning $\Delta$ of $Q_2$ (a two-dimensional chevron scan). The resonance condition of Eq.~\eqref{eq:resonance} appears as a bright central ridge of maximum population transfer to $|020\rangle$. Zero detuning ($\Delta = 0$) gives the maximal Rabi amplitude and defines the optimal operating point.
		(b) Coherent population exchange at the resonant point fixed in (a), showing the time evolution of the populations in $|101\rangle$ (blue) and $|020\rangle$ (red). The system undergoes a full Rabi cycle with period $t_g = 105\,\mathrm{ns}$, corresponding to a $2\pi$ rotation. The vertical dashed line indicates the gate time $t_g$, when the state returns to $|101\rangle$ having acquired the target $\pi$ geometric phase.
	}
\end{figure}

This section details the implementation protocol and validates its experimental feasibility through open-system dynamics simulations. We use the effective Hamiltonian of Eq.~\eqref{eq:H2} together with realistic device parameters and decoherence channels from Appendix~B, confirming both the robust activation of the three-body interaction and the suppression of associated error mechanisms.

The control strategy employs a hierarchical calibration. We first characterize the effective coupling strengths $\tilde{g}_{ij}$ as functions of the coupler frequencies (the spectroscopic mapping is provided in Appendix~B). Experimentally, these frequencies are set deterministically via external flux biases, utilizing the flux-tunable transmon SQUID relation. Based on this characterization, we define a global idle configuration where all residual inter-qubit couplings are suppressed to negligible levels. The system remains at this isolation point during idle periods. From this baseline, shaped frequency pulses—corresponding to flux modulations—are applied to dynamically bring the system into the target resonance regime required for gate operations.

The gate is executed via the two-stage protocol illustrated in Fig.~\ref{fig:sequence}(a), with the complete time-domain control waveforms shown in Fig.~\ref{fig:sequence}(b). Starting from the idle point, the sequence proceeds in two steps. First, synchronized flux pulses drive the central qubit $Q_2$ and the couplers during geometric phase imprinting. These pulses induce a cyclic trajectory between $|101\rangle$ and $|020\rangle$, accumulating the target conditional geometric phase. Second, the active cancellation of residual phases stage addresses deterministic dynamical phases and residual $ZZ$ and $ZZZ$ interactions induced by the fast control pulses. Here, single-qubit phases are compensated via virtual frame updates, and the multi‑body terms are actively canceled. This correction ensures that the net evolution implements a high‑fidelity, diagonal CCZ unitary.

The following subsections detail the calibration physics for these two stages. We then extend the protocol to a Continuous \texorpdfstring{$\mathrm{CCPhase}(\theta)$}{CCPhase} gate set.

\subsection{Calibration of the Three-Body Interaction}

We begin by calibrating the resonant three-body interaction. Our simulation applies synchronized control pulses that dynamically tune the qubit and coupler frequencies from the global idle point into the three-body resonance regime. All control waveforms use a flat-top envelope with $2\,\mathrm{ns}$ cosine-smoothed rising and falling edges. This smooth temporal profile minimizes the spectral bandwidth of the control fields, suppressing leakage to spectator modes and reducing non-adiabatic transitions. Recognizing that generating $2\,\mathrm{ns}$ edges may be constrained by the finite bandwidth of arbitrary waveform generators in certain experimental setups~\cite{RevModPhys.93.025005}, we have systematically evaluated the protocol's performance under longer ramp times. As detailed in Appendix~B, our scheme maintains robust, high-fidelity operations even with significantly extended pulse edges, confirming its broad experimental compatibility.

To locate the optimal operating point, we perform a two-dimensional parameter sweep. We vary both the amplitude of the $Q_2$ pulse—setting its frequency detuning explicitly as $\Delta = (2\tilde{\omega}_2 + \alpha_2) - (\tilde{\omega}_1 + \tilde{\omega}_3)$ from the exact two-photon resonance—and the pulse duration, while the couplers are biased to provide sufficient interaction strength. This simulated sweep is analogous to standard chevron spectroscopy~\cite{PhysRevLett.107.080502}. As shown in Fig.~\ref{fig:calibration}(a), the resonance condition of $\omega_1 + \omega_3 \approx 2\omega_2(t) + \alpha_2$ manifests as a central ridge where population transfer between $|101\rangle$ and $|020\rangle$ is maximized.

With the pulse amplitude fixed at this resonance, we simulate the resulting time-domain Rabi oscillations (Fig.~\ref{fig:calibration}(b)). The system undergoes the cyclic evolution $|101\rangle \to i|020\rangle \to -|101\rangle$. From these oscillations, we extract the precise gate time $t_g$ required for a full $2\pi$ rotation. At $t = t_g$, the population fully returns to $|101\rangle$, having accumulated the target conditional geometric phase of $\pi$.

\subsection{Active Cancellation of Residual Phases}

The driven evolution $|101\rangle \leftrightarrow |020\rangle$ successfully imprints the target conditional geometric phase, but the fast non-adiabatic control also activates residual diagonal interactions. Left uncorrected, these parasitic terms degrade the gate fidelity by accumulating spurious dynamical phases. We model these error channels systematically with the effective Hamiltonian
\begin{equation}
	\label{eq:Vzz}	
	\tilde{H}_{\mathrm{err}} = \sum_{i<j} \zeta_{ij} \hat{n}_i \hat{n}_j + \zeta_{123} \hat{n}_1 \hat{n}_2 \hat{n}_3,
\end{equation}
where $\hat{n}_k = \hat{a}^\dagger_k \hat{a}_k$ is the number operator, $\zeta_{ij}$ and $\zeta_{123}$ are the static coupling coefficients for the residual two-body ($ZZ$) and three-body ($ZZZ$) interactions, respectively.

After the geometric phase imprinting stage, we first remove the deterministic single-qubit dynamical phases $\phi_s$ via instantaneous virtual-Z (VZ) frame updates~\cite{PhysRevA.96.022330}. The remaining multi-body dynamical phases are directly determined by the time integral of the residual coupling strengths over the gate duration $t_g$ (i.e., $\varphi_{ij} = \int_0^{t_g} \zeta_{ij}(t) dt$ and $\varphi_{123} = \int_0^{t_g} \zeta_{123}(t) dt$). Consequently, the remaining unitary $U$ is diagonal in the computational basis:
\begin{equation}
	U=\begin{pmatrix}
		1 & & & & & & & \\
		& 1 & & & & & & \\
		& & 1 & & & & & \\
		& & & e^{-\mathrm{i}\varphi_{|011\rangle}} & & & & \\
		& & & & 1 & & & \\
		& & & & & e^{-\mathrm{i}\varphi_{|101\rangle}} & & \\
		& & & & & & e^{-\mathrm{i}\varphi_{|110\rangle}} & \\
		& & & & & & & e^{-\mathrm{i}\varphi_{|111\rangle}}
	\end{pmatrix}.
\end{equation}

The accumulated phases on the respective states are identified as $\varphi_{|011\rangle} = \varphi_{23}$, $\varphi_{|110\rangle} = \varphi_{12}$, and $\varphi_{|111\rangle} = \varphi_{12} + \varphi_{23} + \varphi_{13} + \varphi_{123}$. Notably, the total phase on the target state is $\varphi_{|101\rangle} = \varphi_{13} + \varphi_{\mathrm{geo}}$. We emphasize that the weak next-nearest-neighbor dynamical phase $\varphi_{13}$ arises entirely from the higher-order effective $ZZ$ cross-talk ($\zeta_{13}$) mediated by the central modes during the driven evolution.

To recover the canonical diagonal form of the CCZ gate, we employ a targeted error-suppression strategy. We first address the dominant nearest-neighbor errors $\varphi_{12}$ and $\varphi_{23}$. These are characterized via Ramsey-type interferometry and actively canceled by applying short corrective CPhase pulses (pulse details are given in Appendix~B). Our simulations determine the required pulse durations to be $\tau_{12} \approx 24\,\mathrm{ns}$ and $\tau_{23} \approx 26\,\mathrm{ns}$. These pulses utilize the standard $|11\rangle \leftrightarrow |02\rangle$ resonance of the respective qubit pairs and are included in the total gate duration.

We then treat the non-nearest-neighbor interaction $\varphi_{13}$. Owing to the structural suppression of direct coupling between $Q_1$ and $Q_3$ in our linear chain, this term is only a small perturbation. Rather than adding an extra cancellation pulse, we absorb $\varphi_{13}$ by fine-tuning the control parameters for the geometric phase accumulation to satisfy $\varphi_{\mathrm{geo}} = \pi - \varphi_{13}$. With this calibration, the effective propagator converges to
\begin{equation}
	\label{eq:U_final}
	U_{\mathrm{eff}} \approx \begin{pmatrix}
		1 & & & & & & & \\
		& 1 & & & & & & \\
		& & 1 & & & & & \\
		& & & 1 & & & & \\
		& & & & 1 & & & \\
		& & & & & -1 & & \\
		& & & & & & 1 & \\
		& & & & & & & 1
	\end{pmatrix}.
\end{equation}

The resulting evolution therefore faithfully implements the canonical CCZ gate, confirming the feasibility of our high-fidelity protocol.

\begin{figure}[t]
	\centering
	\includegraphics[width=\linewidth]{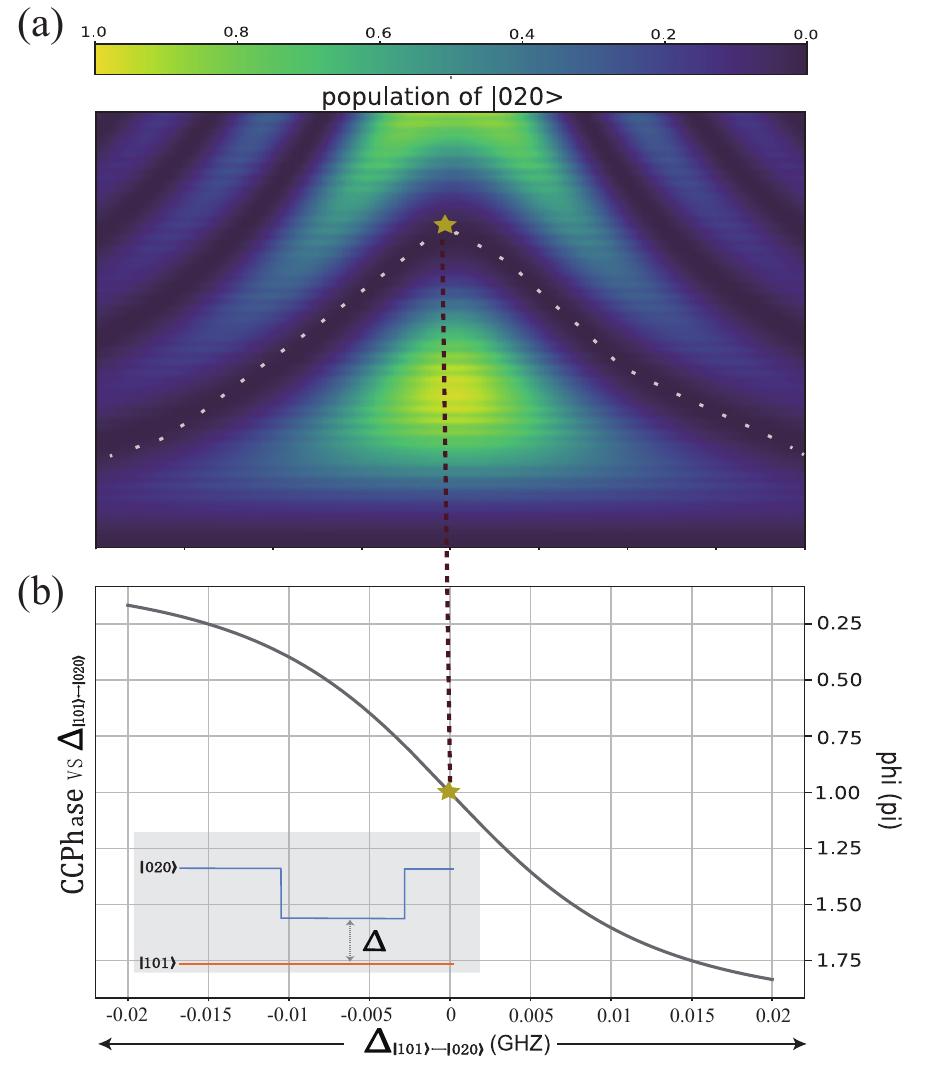}
	\caption{\label{fig:ccphase}
		(a) Parameter landscape for continuous phase control, showing the final population of the auxiliary state $|020\rangle$ as a function of interaction duration $\tau$ (vertical axis) and frequency detuning $\Delta$ (horizontal axis). The white dashed line indicates the closed trajectory that satisfies the cyclic evolution condition—the $(\Delta, \tau)$ combinations for which the population fully returns to the computational subspace.
		(b) Accumulated geometric phase along the closed trajectory, plotted as the conditional phase $\theta$ versus detuning $\Delta$. Tuning $\Delta$ allows $\theta$ to be varied continuously across the full range $[0, 2\pi)$, demonstrating the programmable nature of the geometric phase. The star marks the CCZ operating point ($\theta = \pi$ at $\Delta = 0$), which serves as a benchmark for the entire gate family (cf. Fig.~\ref{fig:calibration}).
	}
\end{figure}

\subsection{Extension to a Continuous \texorpdfstring{$\mathrm{CCPhase}(\theta)$}{CCPhase} Gate Set}
\label{subsec:CCPhase_theta}

The geometric nature of our protocol allows it to be extended beyond the discrete $\pi$ phase~\cite{PhysRevLett.58.1593}, enabling the implementation of an arbitrary conditional phase $\theta$. Continuous phase control is achieved by modifying the evolution trajectory within the auxiliary manifold. Specifically, we jointly tune the frequency detuning $\Delta$ (from the exact three-body resonance) and the interaction duration $\tau$; these together determine the effective solid angle enclosed by the state path and hence the acquired geometric phase.

To maintain a cyclic evolution—where the population fully returns from $|020\rangle$ to the computational subspace—the duration $\tau$ must be recalibrated precisely for each detuning $\Delta$. High-fidelity operations are therefore confined to specific closed-loop contours in the control landscape where leakage is minimized. Along these contours, the accumulated phase becomes a well-defined function $\theta(\Delta, \tau)$. With active cancellation of residual phases (Sec.~\ref{Sec3}B) appropriately applied, the effective unitary takes generic diagonal form:
\begin{equation}
	\label{eq:Uccp}
	U(\theta) \approx \begin{pmatrix}
		1 & & & & & & & \\
		& 1 & & & & & & \\
		& & 1 & & & & & \\
		& & & 1 & & & & \\
		& & & & 1 & & & \\
		& & & & & e^{-\mathrm{i}\theta} & & \\
		& & & & & & 1 & \\
		& & & & & & & 1
	\end{pmatrix}.
\end{equation}

Figure~\ref{fig:ccphase} maps the control parameters needed to span the full phase range $\theta \in [0, 2\pi)$. The CCZ gate ($\theta=\pi$) requires the longest interaction time and is most sensitive to decoherence; its successful implementation therefore validates the entire gate family. This parametric control enables on‑demand synthesis of programmable three‑qubit entanglement, providing a versatile resource for variational quantum algorithms and other protocols that require continuously tunable multi‑qubit interactions.

\section{Gate Performance and Robustness Analysis}
\label{Sec4}

We assess the experimental viability of our protocol by simulating open‑system dynamics governed by the Lindblad master equation. The simulations consider realistic device parameters matching state‑of‑the‑art transmon processors (see Appendix~B for details), with base coherence times $T_1 = 100\,\mu\mathrm{s}$ and $T_2^* = 30\,\mu\mathrm{s}$. To account for the accelerated relaxation of higher energy levels, we scale the decay rates according to harmonic scaling ($\Gamma_{n\to n-1} \approx n\,\Gamma_{1\to0}$), giving a conservative lifetime $T_1^{(|2\rangle)} \approx T_1/2 = 50\,\mu\mathrm{s}$~\cite{PhysRevApplied.23.034046} for the second excited state and preventing an overestimation of fidelity during its transient population.

We evaluate the performance in three stages. First, we report the average gate fidelity and visualize the realized unitary to verify the precise imprinting of the geometric phase. Second, we perform a detailed error‑budget analysis, decomposing the total infidelity into its dominant sources: incoherent errors (from finite $T_1$ and $T_2^*$), leakage outside the computational subspace, and residual coherent control errors. This decomposition is obtained by comparing simulations with all dissipative channels against those without relaxation and dephasing. Finally, we examine the scheme’s practical robustness by mapping the fidelity landscape under static parameter drifts and dynamic control noise.

\subsection{Gate Fidelity and Unitary Evolution}

We use the standard average gate fidelity~\cite{PEDERSEN200747} to quantify the gate performance. For open‑system dynamics described by the Lindblad master equation, the evolution is a quantum channel (superoperator) $\mathcal{E}$. The fidelity is evaluated via the process‑fidelity correspondence
\begin{equation}
	\begin{aligned}
		\mathcal{F}_{\mathrm{avg}} &= \frac{d \mathcal{F}_{\mathrm{pro}} + 1}{d+1}, \\
		\text{with}\quad \mathcal{F}_{\mathrm{pro}} &= \frac{1}{d^2}\mathrm{Tr}\bigl(\mathcal{S}_{\mathrm{ideal}}^\dagger \mathcal{S}_{\mathrm{real}}\bigr),
	\end{aligned}
\end{equation}
where $d=8$ is the Hilbert-space dimension of the three-qubit computational basis. Here, $\mathcal{S}_{\mathrm{real}}$ is the superoperator representation of the simulated quantum channel $\mathcal{E}$, and $\mathcal{S}_{\mathrm{ideal}}$ is the superoperator representation of the ideal CCZ unitary operator.

With the device parameters listed in Appendix~B and the optimized control sequence, we obtain an average gate fidelity $\mathcal{F}_{\mathrm{avg}} > 99\%$ within a total gate duration $t_{\mathrm{gate}} \approx 165\,\mathrm{ns}$. As shown in the control waveform (Fig.~\ref{fig:sequence}(b)), this duration includes the geometric‑phase accumulation ($\tau_{\mathrm{geo}} \approx 105\,\mathrm{ns}$), two active‑cancellation pulses ($\tau_{12} \approx 24\,\mathrm{ns}$, $\tau_{23} \approx 26\,\mathrm{ns}$), and two fixed $5\,\mathrm{ns}$ buffer intervals that isolate the operational stages and suppress transient crosstalk.

\begin{figure}[t]
	\centering
	\includegraphics[width=\linewidth]{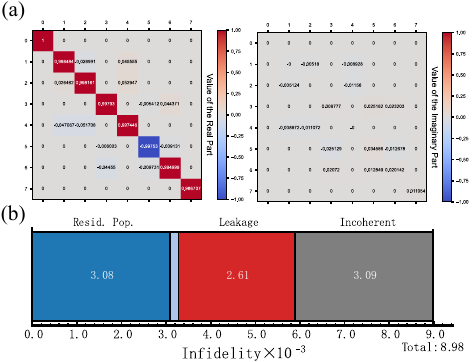}
	\caption{
		(a) Real (left) and imaginary (right) parts of the realized three-qubit unitary in the computational basis $\{|000\rangle, |001\rangle, \dots, |111\rangle\}$. The real part exhibits the target $\pi$ phase shift on state $|101\rangle$ (diagonal element $\approx -1$), confirming the accumulation of the conditional geometric phase. The near-zero imaginary part validates the suppression of off-diagonal coherent errors by the active cancellation protocol.
		(b) Error-budget decomposition for the $t_{\mathrm{gate}} = 165\,\mathrm{ns}$ gate. The total infidelity $1-\mathcal{F}_{\mathrm{avg}} = 8.98 \times 10^{-3}$ ($\mathcal{F}_{\mathrm{avg}} > 99\%$) separates into coherent errors (blue bars), leakage outside the computational subspace ($n \ge 3$, red), and incoherent errors from finite $T_1$ and $T_2^*$ (grey). The coherent contribution splits further: residual non‑adiabatic population in $|011\rangle/|110\rangle$ (darker blue, $3.08\times10^{-3}$) and residual phase miscalibration (lighter blue, $0.2\times10^{-3}$), as detailed in Sec.~\ref{Sec4}. The comparable scale of coherent and incoherent errors indicates that the gate is limited by both control imperfections and decoherence under the given coherence times.
	}
	\label{Fig6}
\end{figure}

To isolate and visualize the coherent logical action of the gate, we compute the unitary evolution operator via lossless Schrödinger dynamics. This simulation uses an extended Hilbert space, truncating each transmon to its lowest 10 levels to explicitly capture possible leakage to high‑energy states. The infidelity extracted from this full‑scale simulation closely matches the prediction of the effective qutrit model ($N=3$) presented in Sec.~\ref{Sec2}, quantitatively confirming that leakage to higher manifolds is negligible. The resulting propagator, projected into the eight‑dimensional computational subspace, is plotted in Fig.~\ref{Fig6}(a). While the average fidelity $\mathcal{F}_{\mathrm{avg}}$ reported above incorporates full open‑system decoherence, this unitary representation directly confirms the coherent performance: the precise accumulation of a $\pi$ geometric phase on $|101\rangle$ and the effective cancellation of residual diagonal phase errors by the active error‑suppression stage.

\subsection{Error Budget Breakdown}
\label{subsec:error_budget}

To identify the physical mechanisms limiting gate performance, we decompose the total infidelity ($1 - \mathcal{F}_{\mathrm{avg}} \approx 8.98 \times 10^{-3}$) into specific error channels. This breakdown is visualized in Fig.~\ref{Fig6}(b) and quantified in Table~\ref{tab:error_budget}. Following the mechanistic analysis in Appendix~C, we group the errors into three categories: (i) \textit{coherent errors}, including residual populations and phase miscalibrations; (ii) \textit{incoherent errors} from relaxation and dephasing; and (iii) \textit{leakage errors} to levels outside the computational subspace ($n \ge 3$).

\begin{table}[b]
	\centering
	\caption{Error budget of the realized CCZ gate ($t_\mathrm{gate} \approx 165\,\mathrm{ns}$; $\mathcal{F}_\mathrm{avg} > 99.1\%$). Values are extracted as described in Appendix~C.}
	\label{tab:error_budget}
	\small
	\setlength{\tabcolsep}{3pt} 
	\renewcommand{\arraystretch}{1.3}
	\begin{tabular}{lc}
		\hline \hline
		\textbf{Error Channel} & \textbf{Infidelity} ($\times 10^{-3}$) \\
		\hline
		\multicolumn{2}{l}{\textit{Coherent Control Errors}} \\
		\hspace{3mm}Residual pop. (mainly in $|110\rangle/|011\rangle$) & $3.08$ \\
		\hspace{3mm}Phase miscalibration & $0.20$ \\
		\hline
		\multicolumn{2}{l}{\textit{Incoherent Errors}} \\
		\hspace{3mm}Relaxation ($T_1$) \& dephasing ($T_2^*$) & $3.09$ \\
		\hline
		\multicolumn{2}{l}{\textit{Leakage Errors}} \\
		\hspace{3mm}High-level leakage ($n \ge 3$) & $2.61$ \\
		\hline
		\textbf{Total Infidelity} & $\mathbf{8.98}$ \\
		\hline \hline
	\end{tabular}
\end{table}

We compute these components through differential simulations, comparing full master-equation dynamics with subspace-projected and lossless evolution. As Table~\ref{tab:error_budget} shows, the largest coherent error ($3.08 \times 10^{-3}$) arises from \textit{residual diabatic population} left in the intermediate states $|011\rangle$ and $|110\rangle$. This error results from the finite spectral bandwidth of the fast control pulses and reflects the inherent trade-off between gate speed ($165\,\mathrm{ns}$) and adiabaticity in our scheme. In contrast, the \textit{phase miscalibration} error is only $0.20 \times 10^{-3}$, confirming the effectiveness of the active cancellation pulses.

It is important to distinguish this residual population—which stays within the qutrit subspace—from \textit{true leakage} to non‑computational levels. To validate the effective qutrit model of Sec.~\ref{Sec2}, we explicitly extract the population transferred to $n \ge 3$ manifolds from full-system dynamics. This leakage contributes $2.61 \times 10^{-3}$ to the infidelity. Although measurable—a consequence of the aggressive gate timing—it confirms that the pulse shaping largely confines the dynamics to the intended subspace, preventing leakage from dominating the error budget.

The total incoherent error is $3.09 \times 10^{-3}$, which includes the accelerated decay of the auxiliary state $|020\rangle$ (with $T_1^{(|2\rangle)} \approx T_1/2$). The fact that the incoherent contribution is comparable to the combined coherent‑control and leakage errors indicates that the present implementation balances operation speed against decoherence. Nevertheless, because the sum of coherent and leakage errors still exceeds the incoherent floor, the protocol remains partly \textit{control‑limited}~\cite{PhysRevA.93.012301}. This suggests that further suppression of leakage and non‑adiabatic transitions—for instance, via advanced pulse‑shaping techniques such as GRAPE~\cite{KHANEJA2005296} or DRAG~\cite{PhysRevLett.103.110501}—could improve fidelity before reaching the hard limit set by $T_1$.

\begin{figure}[t]
	\centering
	\includegraphics[width=\linewidth]{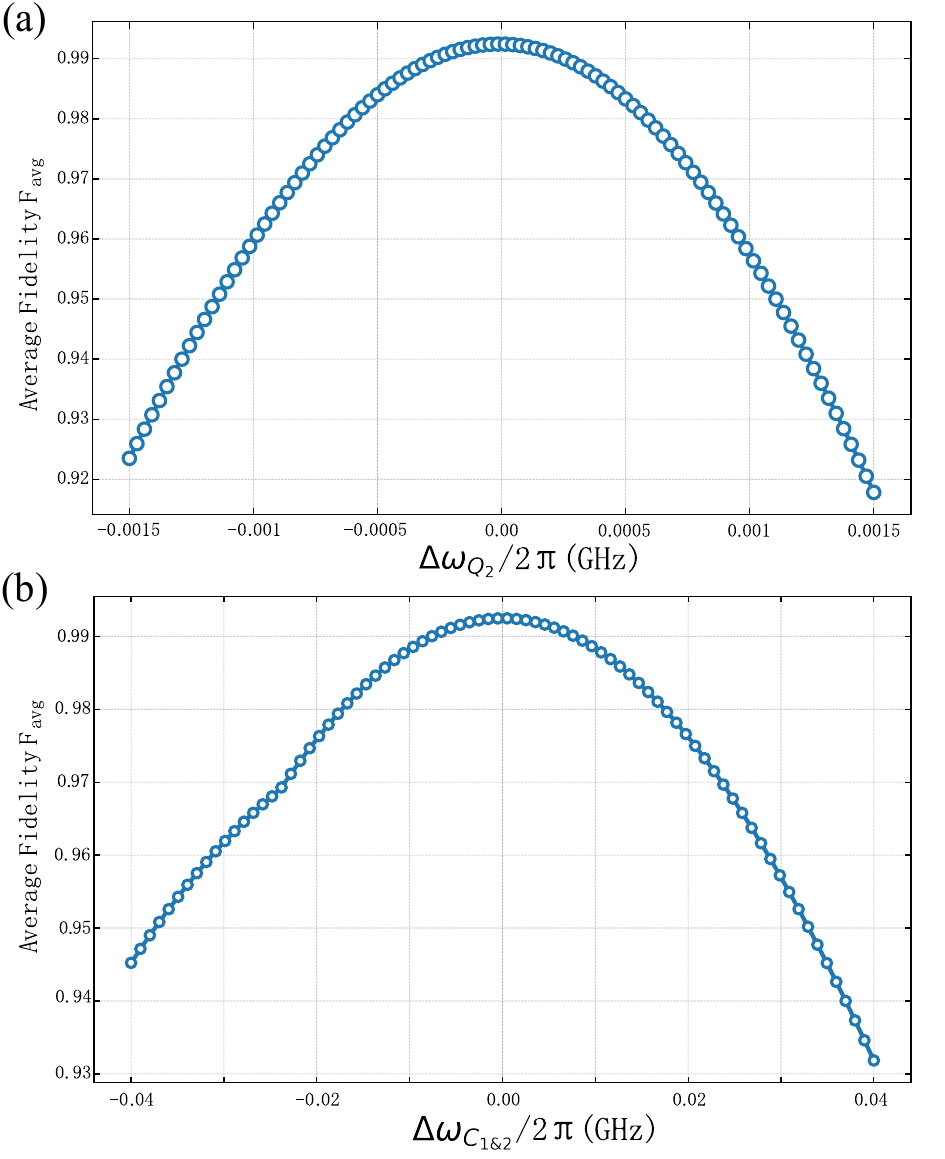}
	\caption{
		(a) Sensitivity of the average gate fidelity $\mathcal{F}_{\mathrm{avg}}$ to the frequency detuning of the central qubit, defined as $\Delta\omega_{Q_2} = \tilde{\omega}_2 - \tilde{\omega}_2^{\mathrm{opt}}$ (where $\tilde{\omega}_2^{\mathrm{opt}}$ is the ideal frequency satisfying the exact resonance condition). The bell-shaped response maintains $\mathcal{F}_{\mathrm{avg}} > 92\%$ over $\pm 1.5\,\mathrm{MHz}$, showing that the geometric-phase accumulation is tolerant to deviations from the exact resonance. 
		(b) Fidelity under a common-mode frequency detuning $\Delta\omega_{C_{1\&2}} = \tilde{\omega}_{C_{1,2}} - \tilde{\omega}_{C_{1,2}}^{\mathrm{opt}}$ of both tunable couplers over $\pm 40\,\mathrm{MHz}$ relative to their designated operating points. This scan varies the effective couplings $|\tilde{g}_{12}|$ from $17.57\,\mathrm{MHz}$ to $15.92\,\mathrm{MHz}$ and $|\tilde{g}_{23}|$ from $15.58\,\mathrm{MHz}$ to $14.16\,\mathrm{MHz}$, thereby modulating the three-body coupling $J$ (see Eq.~\ref{eq:J_eff}). Despite this substantial variation, $\mathcal{F}_{\mathrm{avg}}$ remains above $93\%$ throughout, confirming the robustness of the mechanism against typical coupler-bias fluctuations.
	}
	\label{fig:robustness}
\end{figure}

\subsection{Robustness and Comparative Advantage}
\label{subsec:robustness}

To quantify the robustness of our geometric protocol, we map the gate fidelity against deviations in two critical control parameters: (i) the frequency of the central qubit $Q_2$, which sets the three‑body resonance condition, and (ii) the frequencies of the tunable couplers, which determine the effective nearest‑neighbor couplings $\tilde{g}_{ij}$.

As shown in Fig.~\ref{fig:robustness}, the gate retains high fidelity over a broad operational window. For the central‑qubit frequency, the fidelity remains above $92\%$ across a detuning range of $\pm 1.5$~MHz (Fig.~\ref{fig:robustness}(a)), demonstrating tolerance to deviations from the exact resonance condition $\tilde{\omega}_1 + \tilde{\omega}_3 \approx 2\tilde{\omega}_2 + \alpha_2$. For the couplers, a common‑mode drift of $\pm 40$~MHz changes the effective couplings $|\tilde{g}_{12}|$ from $17.57$~MHz to $15.92$~MHz and $|\tilde{g}_{23}|$ from $15.58$~MHz to $14.16$~MHz, thereby modulating the three‑body coupling $J$ (see Eq.~\ref{eq:J_eff}). Despite this substantial variation, fidelity stays above $93\%$ throughout the range (Fig.~\ref{fig:robustness}(b)), confirming the robustness of the geometric‑phase mechanism against typical coupler‑bias fluctuations. The smooth, featureless response in both scans indicates that the shaped pulses avoid accidental resonances with spectator modes such as $|210\rangle$, which would otherwise produce narrow dips in the fidelity landscape. This stability, together with the active suppression of leakage channels via pulse shaping (Sec.~\ref{subsec:error_budget}), enables a reliable operation regime that reduces dependency on exhaustive heuristic calibration.

The practical advantage of our native gate is underscored by direct comparison with both standard decomposition and the state‑of‑the‑art experimental implementation. Recent experimental characterization on similar tunable‑coupler hardware reveals that constructing a CCZ gate via standard decomposition requires a sequence of 8 CNOT gates interspersed with single‑qubit rotations, extending the total execution time to $\sim 640\,\mathrm{ns}$ and incurring significant cumulative coherent and leakage errors~\cite{lvb9-pfr3}. In contrast, our direct implementation realizes the CCZ gate natively within $165\,\mathrm{ns}$, achieving a $3.9\times$ speedup over the decomposed sequence and drastically reducing the exposure to decoherence. Under realistic error models, this temporal compression directly translates to a higher achievable fidelity.

The benefit of a hardware-efficient native design is further highlighted by comparing with the experimental demonstration of a direct CCZ gate in the same work~\cite{lvb9-pfr3}. While their result validates the experimental viability of the direct-gate paradigm—achieving a process fidelity of $93.54\%$ in $256\,\mathrm{ns}$—their gate dynamics fundamentally rely on accessing the more decoherence-susceptible three-excitation manifold (e.g., the $|111\rangle$ state). In explicit contrast, our protocol strictly confines the interaction to the two-excitation manifold, centered on the resonant $|101\rangle \leftrightarrow |020\rangle$ transition. This mechanism inherently compresses the operation, achieving a gate duration of $165\,\mathrm{ns}$—a reduction of $\sim 35\%$ compared to the experimental benchmark. Moreover, our simulation predicts an average gate fidelity exceeding $99\%$, indicating a clear path to higher performance. This improvement stems from the intrinsic efficiency of the direct three‑body interaction, further enhanced by optimized pulse shaping and active cancellation of residual phases (Sec.~\ref{Sec3}) to suppress leakage and coherent errors. Therefore, our work provides a refined theoretical blueprint where the core resonant mechanism, supported by practical error suppression, charts a path for future implementations toward higher fidelities.

In summary, by combining intrinsic parametric robustness with a compact, fast implementation, our protocol provides a hardware‑efficient route to high‑fidelity three‑qubit logic. It sidesteps the overhead and error accumulation of multi‑gate decompositions while offering a clear pathway to outperform current experimental benchmarks, thereby contributing a scalable primitive for near‑term quantum algorithms and future fault‑tolerant architectures.

\section{Conclusion}
\label{Sec5}

We have demonstrated a hardware-efficient protocol for implementing a native three-qubit CCZ gate on a tunable-coupler superconducting processor. By engineering a resonant three-body interaction via a time-frequency correlated virtual transition, the protocol directly realizes the requisite logic operation in a single step, circumventing the circuit depth and error accumulation inherent in gate-decomposition approaches. The same physical mechanism inherently supports a continuous $\mathrm{CCPhase}(\theta)$ gate set, enabling programmable geometric phases through control parameter variation. 

Open-system simulations based on state-of-the-art device parameters show an average gate fidelity exceeding 99\% within $165\,\mathrm{ns}$ for the CCZ gate. This represents a $\sim 35\%$ reduction in gate duration and a significant fidelity improvement over recent experimental benchmarks for direct CCZ gates on similar hardware. Moreover, the protocol exhibits intrinsic parametric robustness, maintaining high fidelity over realistic operational windows (Sec.~\ref{subsec:robustness}), which simplifies experimental calibration.

A detailed error-budget analysis confirms the scheme's experimental feasibility, revealing that the present performance is limited primarily by coherent control errors—chiefly residual non-adiabatic population—rather than by fundamental decoherence bounds ($T_1$, $T_2^*$). Operating in this control-limited regime implies that fidelity can be further improved with advanced pulse-shaping techniques. Crucially, the protocol relies entirely on existing tunable-coupler hardware and requires no specialized components.

By combining parametric robustness with a compact, fast implementation, our work provides a hardware-efficient route to programmable three-qubit entanglement. This protocol sidesteps the overhead and error accumulation of multi‑gate decompositions while offering a clear pathway to outperform current experimental benchmarks. It thus establishes a high-fidelity primitive that reduces the resource overhead for non-Clifford operations, contributing to both near‑term quantum algorithms and the long‑term development of fault-tolerant quantum architectures.

\section{Acknowledgement}
We would like to thank Wen-Hui Huang and Zi-Yu Tao from the Shenzhen International Quantum Academy, as well as Meng-Ru Yun from Henan Academy of Sciences, for their technical support to this work. This work was supported by the National Key R\&D Program of China (No. 2024YFB4504101), the National Key R$\&$D Program of China (No. 2024YFB4504600) and the National Natural Science Foundation of China (No. 62572482).

\appendix
\setcounter{figure}{0}
\setcounter{table}{0}
\section{Detailed Derivation of the Effective Three-Body Interaction}
\label{app:derivation}

We present a detailed derivation of the effective three-qubit Hamiltonian and the resulting three-body coupling strength $J$.

\subsection{Effective Three-Qubit Hamiltonian via Schrieffer-Wolff Transformation}
\label{subsec:5to3}

We derive the effective three-qubit Hamiltonian starting from the full five-qubit model given in Eq.~\eqref{eq:hami} of the main text:
\begin{equation}
	\begin{aligned}
		\hat{H} &= \sum_{k \in \{1,2,3,c_1,c_2\}} \Bigl( \omega_k \hat{a}^\dagger_k \hat{a}_k + \frac{\alpha_k}{2} \hat{a}^\dagger_k \hat{a}^\dagger_k \hat{a}_k \hat{a}_k \Bigr) \\
		&\quad + \sum_{\langle i,j \rangle} g_{ij} (\hat{a}^\dagger_i - \hat{a}_i)(\hat{a}_j - \hat{a}^\dagger_j).
	\end{aligned}
\end{equation}

It is separated into an unperturbed part $H_0$ and a perturbation $V$:
\begin{equation}
	\begin{aligned}
		\hat{H}_0 &= \sum_{k} \Bigl( \omega_k \hat{a}^\dagger_k \hat{a}_k + \frac{\alpha_k}{2} \hat{a}^\dagger_k \hat{a}^\dagger_k \hat{a}_k \hat{a}_k \Bigr), \\
		\hat{V} &= \sum_{\langle i,j \rangle} g_{ij} \bigl( \hat{a}^\dagger_i \hat{a}_j + \hat{a}_i \hat{a}^\dagger_j - \hat{a}^\dagger_i \hat{a}^\dagger_j - \hat{a}_i \hat{a}_j \bigr),
	\end{aligned}
\end{equation}
with the capacitive coupling expanded into both rotating and counter-rotating terms.

In the dispersive regime ($|g_{ij}| \ll |\omega_i \pm \omega_j|$), we apply a Schrieffer-Wolff (SW) transformation $U = e^{S}$ to decouple the couplers from the qubits. The generator $S$ is anti-Hermitian and satisfies $[H_0, S] = -V$. We construct $S = \sum_{\langle i,j \rangle} S_{ij}$, where each component accounts for the corresponding pairwise interaction.

Because the transmon anharmonicity is substantially larger than the inter-qubit coupling strengths ($|\alpha_k| \gg |g_{ij}|$), its correction to the off-diagonal dispersive coupling is heavily suppressed. Therefore, we safely neglect the influence of $\alpha_k$ on the transformation generator to leading order, and use the simplified commutation relations $[H_0, a^\dagger_i a_j] \approx \Delta_{ij} a^\dagger_i a_j$ and $[H_0, a^\dagger_i a^\dagger_j] \approx \Sigma_{ij} a^\dagger_i a^\dagger_j$ to find:
\begin{equation}
	\hat{S}_{ij} = \frac{g_{ij}}{\omega_i - \omega_j}(\hat{a}_i \hat{a}^\dagger_j - \hat{a}^\dagger_i \hat{a}_j) + \frac{g_{ij}}{\omega_i + \omega_j}(\hat{a}_i \hat{a}_j - \hat{a}^\dagger_i \hat{a}^\dagger_j).
\end{equation}
The effective Hamiltonian is then obtained via the second-order expansion:
\[
H_{\text{eff}} = e^{S} H e^{-S} \approx H_0 + \frac{1}{2}[S, V].
\]
Tracing out the coupler modes by projecting onto their vacuum states (i.e., $\langle \hat{a}^\dagger_{c} \hat{a}_{c} \rangle = 0$), we obtain the effective three-qubit Hamiltonian:
\begin{equation}
	\begin{aligned}
		\hat{H}_{\mathrm{eff}}^{\mathrm{3q}} = & \sum_{i=1}^{3} \Bigl( \tilde{\omega}_i \hat{a}^\dagger_i \hat{a}_i + \frac{\alpha_i}{2} \hat{a}^\dagger_i \hat{a}^\dagger_i \hat{a}_i \hat{a}_i \Bigr) \\
		& + \tilde{g}_{12} (\hat{a}^\dagger_1 \hat{a}_2 + \hat{a}_1 \hat{a}^\dagger_2) + \tilde{g}_{23} (\hat{a}^\dagger_2 \hat{a}_3 + \hat{a}_2 \hat{a}^\dagger_3) \\
		& + \tilde{g}_{13} (\hat{a}^\dagger_1 \hat{a}_3 + \hat{a}_1 \hat{a}^\dagger_3),
	\end{aligned}
	\label{eq:H_eff_3q_app}
\end{equation}
where the renormalized frequencies include the dispersive Lamb shifts from all connected modes:
\[
\tilde{\omega}_i \approx \omega_i + \sum_{j \neq i} g_{ij}^2 \Bigl( \frac{1}{\omega_i - \omega_j} - \frac{1}{\omega_i + \omega_j} \Bigr).
\]
To second order in the couplings, the effective inter-qubit couplings are given by:
\begin{equation}
	\begin{split}
		\tilde{g}_{ij} \approx g_{ij} + \sum_{k=c_1, c_2} \frac{g_{ik} g_{jk}}{2} 
		&\biggl(\frac{1}{\omega_i - \omega_k} + \frac{1}{\omega_j - \omega_k} \\
		&- \frac{1}{\omega_i + \omega_k} - \frac{1}{\omega_j + \omega_k} \biggr).
	\end{split}
\end{equation}

In our specific architecture ($Q_1$--$C_1$--$Q_2$--$C_2$--$Q_3$), we assume negligible cross-coupling ($g_{1c_2} \approx 0, g_{3c_1} \approx 0$) and negligible direct interaction ($g_{13} \approx 0$). Consequently, the expressions simplify to:
\begin{align}
	\tilde{g}_{12} &\approx g_{12} + \frac{g_{1c_1} g_{2c_1}}{2} \Bigl( \frac{1}{\Delta_{1c_1}} + \frac{1}{\Delta_{2c_1}} - \frac{1}{\Sigma_{1c_1}} - \frac{1}{\Sigma_{2c_1}} \Bigr), \label{eq:gtilde12}\\
	\tilde{g}_{23} &\approx g_{23} + \frac{g_{2c_2} g_{3c_2}}{2} \Bigl( \frac{1}{\Delta_{2c_2}} + \frac{1}{\Delta_{3c_2}} - \frac{1}{\Sigma_{2c_2}} - \frac{1}{\Sigma_{3c_2}} \Bigr), \label{eq:gtilde23}\\
	\tilde{g}_{13} &\approx 0, \label{eq:gtilde13}
\end{align}
where we have defined the detunings $\Delta_{ic} = \omega_i - \omega_c$ and sum frequencies $\Sigma_{ic} = \omega_i + \omega_c$. As noted in Sec.~\ref{Sec2}, $\tilde{g}_{13}$ is omitted in the final model.

\subsection{Derivation of the Three-Body Coupling $J$}
\label{subsec:J_app}

Within the effective three-qubit Hamiltonian $\hat{H}_{\mathrm{eff}}^{\mathrm{3q}}$ [Eq.~\eqref{eq:H_eff_3q_app}], we derive the effective coupling strength $J$ between the computational state $|101\rangle$ and the auxiliary state $|020\rangle$ using second-order perturbation theory. We consider the subspace spanned by: $\{|101\rangle, |011\rangle, |110\rangle, |020\rangle\}$.

The unperturbed energies (diagonal elements) are:
\begin{align}
	E_{101} &= \tilde{\omega}_1 + \tilde{\omega}_3, & E_{020} &= 2\tilde{\omega}_2 + \alpha_2, \notag \\
	E_{011} &= \tilde{\omega}_2 + \tilde{\omega}_3, & E_{110} &= \tilde{\omega}_1 + \tilde{\omega}_2.
\end{align}
The relevant off-diagonal matrix elements are:
\begin{align}
	\langle 011| \hat{H}_{\mathrm{eff}}^{\mathrm{3q}} |101\rangle &= \tilde{g}_{12}, & \langle 020| \hat{H}_{\mathrm{eff}}^{\mathrm{3q}} |011\rangle &= \sqrt{2}\,\tilde{g}_{23}, \notag \\
	\langle 110| \hat{H}_{\mathrm{eff}}^{\mathrm{3q}} |101\rangle &= \tilde{g}_{23}, & \langle 020| \hat{H}_{\mathrm{eff}}^{\mathrm{3q}} |110\rangle &= \sqrt{2}\,\tilde{g}_{12}.
\end{align}

The gate operation relies on the resonance condition $E_{101} \approx E_{020}$, achieved by flux-tuning qubit~2 such that $\tilde{\omega}_1 + \tilde{\omega}_3 \approx 2\tilde{\omega}_2 + \alpha_2$. Under this condition, and provided the intermediate states $|011\rangle$ and $|110\rangle$ remain far off-resonance ($|\tilde{g}_{ij}| \ll |\tilde{\omega}_i - \tilde{\omega}_j|$), the effective coupling $J$ is generated via two virtual pathways:
\begin{equation}
	J \approx \sum_{m \in \{011,110\}} \frac{\langle 020| \hat{H}_{\mathrm{eff}}^{\mathrm{3q}} | m \rangle \langle m | \hat{H}_{\mathrm{eff}}^{\mathrm{3q}} |101\rangle}{E_{101} - E_m}.
\end{equation}

\noindent \textit{Pathway 1 (via $|011\rangle$):}
The energy detuning is $\Delta E_1 = E_{101} - E_{011} = \tilde{\omega}_1 - \tilde{\omega}_2$. The main contribution is:
\[
J_1 = \frac{\sqrt{2}\,\tilde{g}_{23} \tilde{g}_{12}}{\tilde{\omega}_1 - \tilde{\omega}_2}.
\]

\noindent \textit{Pathway 2 (via $|110\rangle$):}
The energy detuning is $\Delta E_2 = E_{101} - E_{110} = \tilde{\omega}_3 - \tilde{\omega}_2$. The main contribution is:
\[
J_2 = \frac{\sqrt{2}\,\tilde{g}_{12} \tilde{g}_{23}}{\tilde{\omega}_3 - \tilde{\omega}_2}.
\]

Summing these contributions, we obtain the total effective coupling:
\begin{equation}
	J = J_1 + J_2 = \sqrt{2}\,\tilde{g}_{12}\tilde{g}_{23} \left( \frac{1}{\tilde{\omega}_1 - \tilde{\omega}_2} + \frac{1}{\tilde{\omega}_3 - \tilde{\omega}_2} \right).
	\label{eq:J_derivation_app}
\end{equation}

This expression confirms that the three-body interaction is activated directly from the renormalized two-body couplings $\tilde{g}_{ij}$ by satisfying the frequency matching condition.

To confirm the consistency of Eq.~\eqref{eq:J_derivation_app} with our simulations, we evaluate \(J\) using the effective qubit frequencies obtained after the Schrieffer-Wolff transformation. These frequencies, which include the static Lamb shifts, are \(\tilde{\omega}_1/2\pi = 5.07\,\mathrm{GHz}\), \(\tilde{\omega}_2/2\pi = 5.16\,\mathrm{GHz}\), and \(\tilde{\omega}_3/2\pi = 4.89\,\mathrm{GHz}\) (rounded to two decimal places). Substituting these values together with \(\tilde{g}_{12}/2\pi = 16.72\,\mathrm{MHz}\) and \(\tilde{g}_{23}/2\pi = 14.85\,\mathrm{MHz}\) into Eq.~\eqref{eq:J_derivation_app} yields a theoretical coupling strength \(|J|/2\pi \approx 5.20\,\mathrm{MHz}\).

In the resonant two-level subspace spanned by \(|101\rangle\) and \(|020\rangle\), the effective Hamiltonian reduces to \(\tilde{H}_{\mathrm{eff}} \approx J |101\rangle\langle 020| + \mathrm{H.c.}\), which drives Rabi oscillations between the two states. The full oscillation period \(T_{\mathrm{Rabi}}\) is related to the coupling strength by \(T_{\mathrm{Rabi}} = \pi / |J|\), or equivalently, the Rabi frequency \(\Omega_{\mathrm{Rabi}} = 2|J|\). Independently, we extract the coupling strength from the simulated Rabi oscillations between \(|101\rangle\) and \(|020\rangle\) shown in Fig.~\ref{fig:calibration}(b). The observed oscillation period of \(105\,\mathrm{ns}\) corresponds to a Rabi frequency \(2|J|/2\pi = 1/(105\,\mathrm{ns}) \approx 9.52\,\mathrm{MHz}\), giving \(|J|/2\pi \approx 4.76\,\mathrm{MHz}\). The close agreement (within \(\sim\!8\%\)) between the theoretical and simulated values validates the perturbative derivation of Eq.~\eqref{eq:J_derivation_app}. The small discrepancy can be attributed to higher-order corrections and AC-Stark shifts not included in the second-order Schrieffer-Wolff expansion.

\subsection{Dynamics of the Virtually Populated States and the Gate Mechanism}
\label{subsec:virtual_dynamics}

To elucidate the physics of the correlated virtual process and rigorously derive the effective two-level gate mechanism [as discussed below Eq.~\eqref{eq:Heff}], we analyze the system dynamics within the four-level relevant subspace spanned by $\{|101\rangle, |110\rangle, |011\rangle, |020\rangle\}$. 

\paragraph{Interaction Picture Transformation.---}
Following the Schrieffer-Wolff transformation, the dynamics are governed by the effective three-qubit Hamiltonian $\hat{H}_{\mathrm{eff}}^{\mathrm{3q}} = \hat{H}_{\mathrm{diag}} + \hat{V}$. Restricted to our target subspace, the unperturbed dressed energy part is $\hat{H}_{\mathrm{diag}} = \sum_{k} E_k |k\rangle\langle k|$, where the state energies are $E_{101} = \tilde{\omega}_1 + \tilde{\omega}_3$, $E_{020} = 2\tilde{\omega}_2 + \alpha_2$, $E_{110} = \tilde{\omega}_1 + \tilde{\omega}_2$, and $E_{011} = \tilde{\omega}_2 + \tilde{\omega}_3$. The transverse coupling part is $\hat{V} = \tilde{g}_{12}(\hat{a}_1^\dagger \hat{a}_2 + \mathrm{H.c.}) + \tilde{g}_{23}(\hat{a}_2^\dagger \hat{a}_3 + \mathrm{H.c.})$. 

We transform the system into the interaction picture defined by $U_I(t) = \exp(i \hat{H}_{\mathrm{diag}} t)$. The state vector becomes $|\psi_I(t)\rangle = C_{101}|101\rangle + C_{110}|110\rangle + C_{011}|011\rangle + C_{020}|020\rangle$. During the flat-top duration of the control pulse, the tuned frequency $\tilde{\omega}_2$ is constant, making the energy detunings time-independent. We define the intermediate detunings relative to the computational state as $\Delta_{110} = E_{110} - E_{101} = \tilde{\omega}_2 - \tilde{\omega}_3$ and $\Delta_{011} = E_{011} - E_{101} = \tilde{\omega}_2 - \tilde{\omega}_1$; and relative to the auxiliary state as $\Delta'_{110} = E_{110} - E_{020}$ and $\Delta'_{011} = E_{011} - E_{020}$. The Schrödinger equation $i\partial_t |\psi_I\rangle = \hat{V}_I(t) |\psi_I\rangle$ yields the explicit coupled differential equations governing the entire four-level subspace:
\begin{align}
	i\dot{C}_{101} &= \tilde{g}_{23} e^{-i \Delta_{110} t} C_{110} + \tilde{g}_{12} e^{-i \Delta_{011} t} C_{011}, \label{eq:C101_I}\\
	i\dot{C}_{020} &= \sqrt{2}\tilde{g}_{12} e^{-i \Delta'_{110} t} C_{110} + \sqrt{2}\tilde{g}_{23} e^{-i \Delta'_{011} t} C_{011}, \label{eq:C020_I}\\
	i\dot{C}_{110} &= \tilde{g}_{23} e^{i \Delta_{110} t} C_{101} + \sqrt{2}\tilde{g}_{12} e^{i \Delta'_{110} t} C_{020}, \label{eq:C110_I}\\
	i\dot{C}_{011} &= \tilde{g}_{12} e^{i \Delta_{011} t} C_{101} + \sqrt{2}\tilde{g}_{23} e^{i \Delta'_{011} t} C_{020}, \label{eq:C011_I}
\end{align}
where the factor of $\sqrt{2}$ arises from the bosonic enhancement of the $|1\rangle \rightarrow |2\rangle$ transition in $Q_2$.

\paragraph{Adiabatic Elimination and the Effective Hamiltonian.---}
In the dispersive regime ($|\tilde{g}_{ij}| \ll |\Delta|, |\Delta'|$), the amplitudes $C_{101}$ and $C_{020}$ evolve slowly compared to the rapid oscillations of the phase factors. Integrating Eqs.~\eqref{eq:C110_I} and \eqref{eq:C011_I} over time and discarding the fast-oscillating boundary terms, the instantaneous amplitudes for the intermediate states are:
\begin{align}
	C_{110}(t) &\approx -\frac{\tilde{g}_{23}}{\Delta_{110}} e^{i \Delta_{110} t} C_{101}(t) - \frac{\sqrt{2}\tilde{g}_{12}}{\Delta'_{110}} e^{i \Delta'_{110} t} C_{020}(t), \label{eq:C110_sol}\\
	C_{011}(t) &\approx -\frac{\tilde{g}_{12}}{\Delta_{011}} e^{i \Delta_{011} t} C_{101}(t) - \frac{\sqrt{2}\tilde{g}_{23}}{\Delta'_{011}} e^{i \Delta'_{011} t} C_{020}(t). \label{eq:C011_sol}
\end{align}
These solutions rigorously demonstrate that the virtual states only acquire perturbative transient populations proportional to $(\tilde{g}/\Delta)^2$, validating the time-frequency correlated mechanism.

Substituting these integrated solutions back into the evolution equations for $C_{101}$ and $C_{020}$, we obtain effective coupling terms carrying phase factors like $e^{i(\Delta'_{110} - \Delta_{110})t}$. Notice that $\Delta'_{110} - \Delta_{110} = \Delta'_{011} - \Delta_{011} = E_{101} - E_{020}$. At our chosen operating point, the flat-top pulse brings the system into exact two-photon resonance ($E_{101} = E_{020}$), which immediately enforces $\Delta'_{110} = \Delta_{110}$ and $\Delta'_{011} = \Delta_{011}$. Under this condition, the exponential phase factors for the cross-coupling terms perfectly evaluate to $1$. This second-order elimination reduces the system to an effective Hamiltonian in the $\{|101\rangle, |020\rangle\}$ subspace:
\begin{equation}
	\begin{split}
		\tilde{H}_{\mathrm{eff}, I} &= \delta_{\mathrm{Stark}, 101}|101\rangle\langle 101| + \delta_{\mathrm{Stark}, 020}|020\rangle\langle 020| \\
		&\quad + \left( J |101\rangle\langle 020| + \mathrm{H.c.} \right),
	\end{split}
\end{equation}
where $J = -\sqrt{2}\tilde{g}_{12}\tilde{g}_{23} (1/\Delta_{110} + 1/\Delta_{011})$ is the effective exchange coupling [explicitly derived in Appendix~\ref{subsec:J_app}], and the diagonal terms $\delta_{\mathrm{Stark}}$ represent the state-dependent energy shifts (including AC-Stark shifts and residual ZZ cross-talks) induced by the virtual transitions.

\paragraph{Elimination of Diagonal Terms and Gate Implementation.---}
To achieve a high-fidelity logical gate, the parasitic diagonal perturbations in $\tilde{H}_{\mathrm{eff}, I}$ must be strictly neutralized. In our scheme, the relative dynamically induced shift ($\delta_{\mathrm{Stark}, 020} - \delta_{\mathrm{Stark}, 101}$) is inherently compensated by finely calibrating the flat-top pulse amplitude ($\delta_{\mathrm{pulse}} = 0$) during Stage 1. Furthermore, the global and residual multi-qubit dynamical phases are deterministically zeroed out by the active cancellation protocol in Stage 2. 

With all diagonal terms physically eliminated from the logical evolution, the net dynamics governing the coherent exchange between the computational and auxiliary states is perfectly captured by the pure off-diagonal Hamiltonian:
\begin{equation}
	\label{eq:Heff_app}
	\tilde{H}_{\mathrm{eff}} = J \hat{a}_1 \hat{a}_3 \hat{\Xi}^+ + \mathrm{H.c.}
\end{equation}
This Hamiltonian drives the required cyclic Rabi oscillation $|101\rangle \rightarrow i|020\rangle \rightarrow -|101\rangle$, seamlessly imprinting the geometric phase of $\pi$ for the native CCZ gate.

\begin{figure}[t]
	\centering
	\includegraphics[width=\linewidth]{FigA2.pdf}
	\renewcommand*{\thefigure}{A\arabic{figure}}
	\caption{
		Robustness of the CCZ gate against residual next-nearest-neighbor coupling $g_{13}$. 
		The plot displays the average gate fidelity as a function of the stray capacitive coupling strength $g_{13}/2\pi$. 
		Even in the presence of finite $g_{13}$ arising from realistic hardware imperfections, the fidelity is robustly maintained above $98\%$ after a minor control parameter re-optimization. 
		Physically, this re-optimization effectively absorbs the parasitic dynamical phase $\varphi_{13}$ into the geometric phase accumulation without requiring additional corrective pulses, demonstrating the high resilience of the protocol.
	}
	\label{FigA2} 
\end{figure}

\begin{figure}[t]
	\centering
	\includegraphics[width=\linewidth]{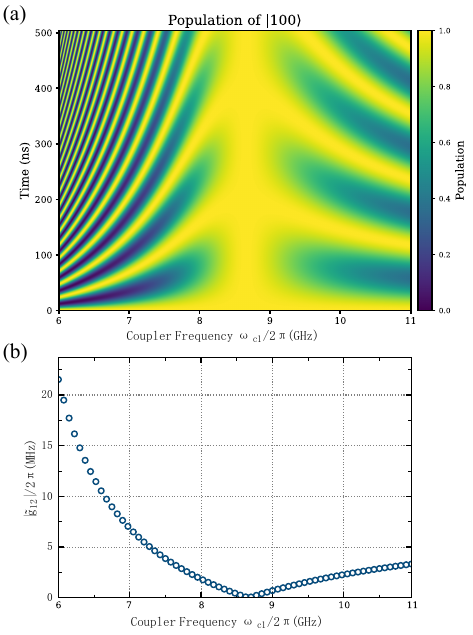}
	\renewcommand*{\thefigure}{A\arabic{figure}}
	\caption{
		(a) Vacuum Rabi oscillations between states $|100\rangle$ and $|010\rangle$ as a function of interaction time and the frequency of coupler $C_1$. 
		(b) Effective coupling strength $\tilde{g}_{12}$ extracted from the oscillation frequency in (a).
	}
	\label{FigA1} 
\end{figure}
	
\begin{table*}[t]
	\centering
	\small
	\setlength{\tabcolsep}{6pt}
	\renewcommand*{\thetable}{A\arabic{table}}
	\caption{Device parameters for the operational stages of the CCZ gate protocol.
	Stage 1 implements the resonant three-body interaction for geometric phase accumulation. Stage 2a and 2b execute the two sequential CPhase gates for active cancellation of residual phases. Note that the qubit ($\omega_i$) and coupler ($\omega_{c_i}$) frequencies listed here represent the bare frequencies before the Schrieffer-Wolff transformation, whereas $\tilde{g}_{ij}$ are the effective (renormalized) nearest-neighbor couplings. Frequencies are in $\mathrm{GHz}/2\pi$, couplings in $\mathrm{MHz}/2\pi$.}
	\label{tab:device_params}
	\begin{tabular}{lcccc}
		\hline\hline
		& \textbf{Idle} & \textbf{Stage 1} & \textbf{Stage 2a} & \textbf{Stage 2b} \\
		& \textbf{(Isolation)} & \textbf{(CCZ Interaction)} & \textbf{(CPhase $Q_1$-$Q_2$)} & \textbf{(CPhase $Q_2$-$Q_3$)} \\
		\hline
		\textit{Qubit frequencies} $\omega_i/2\pi$ \\
		\quad $Q_1$ & 5.089 & 5.089 & 5.089 & 5.089 \\
		\quad $Q_2$ & 5.250 & 5.203 & 5.468 & 5.268 \\
		\quad $Q_3$ & 4.911 & 4.911 & 4.911 & 4.911 \\
		\hline
		\textit{Coupler frequencies} $\omega_{c_i}/2\pi$ \\
		\quad $C_1$ & 8.743 & 6.343 & 6.543 & 8.743 \\
		\quad $C_2$ & 8.768 & 6.368 & 8.768 & 6.418 \\
		\hline
		\textit{Effective couplings} $\tilde{g}_{ij}/2\pi$ \\
		\quad $\tilde{g}_{12}$ & $\sim 0$ & 16.72 & 15.82 & $\sim 0$ \\
		\quad $\tilde{g}_{23}$ & $\sim 0$ & 14.85 & $\sim 0$ & 14.67 \\
		\hline\hline
	\end{tabular}
\end{table*}
\section{Experimental Setup, Calibration Protocols, and Optimization}
\label{AppB}

This appendix details the device parameters, calibration procedures, and fine-tuning underlying our simulations. To rigorously evaluate the gate performance and error budget under realistic experimental conditions (as discussed in Sec.~IV), our numerical simulations employ the Lindblad master equation. The time evolution of the system's density matrix $\rho(t)$ is governed by:
\begin{equation}
	\dot{\rho}(t) = -i [\hat{H}(t), \rho(t)] + \sum_k \gamma_k \left( \hat{L}_k \rho(t) \hat{L}_k^\dagger - \frac{1}{2} \{\hat{L}_k^\dagger \hat{L}_k, \rho(t)\} \right),
\end{equation}
where $\hat{H}(t)$ is the time-dependent effective Hamiltonian of the driven system [explicitly defined as $\hat{H}_{\mathrm{eff}}$ in Eq.~(\ref{eq:H2}) of the main text], $\hat{L}_k$ are the phenomenological collapse operators representing various local dissipation channels, and $\gamma_k$ are the corresponding decay rates.

For our superconducting architecture, we independently model the energy relaxation and pure dephasing for each transmon and coupler mode. 
The energy relaxation channels are described by the annihilation operators $\hat{L}_{1,j} = \hat{a}_j$ with rates $\gamma_{1,j} = 1/T_{1,j}$. Note that this bosonic operator naturally accounts for the enhanced decay rates of higher excited states (e.g., the $|2\rangle$ state relaxes to the $|1\rangle$ state approximately twice as fast as $|1\rangle$ relaxes to $|0\rangle$, which is critical for modeling the decoherence of our auxiliary $|020\rangle$ state). 
The pure dephasing channels are modeled by the photon-number operators $\hat{L}_{\phi,j} = \hat{a}_j^\dagger \hat{a}_j$ with rates $\gamma_{\phi,j} = 1/T_{\phi,j}$. 
The experimentally relevant coherence times $T_{1}$ and $T_2^*$ used in our error budget determine these rates via the standard relation $1/T_2^* = 1/(2T_1) + 1/T_\phi$. The optimized parameters for the idle point and the two main operational stages (and their sub-steps) are summarized in Table~\ref{tab:device_params}.

\paragraph{Device parameters and pulse shaping.}
We simulate a system with typical tunable-coupler transmon parameters: base relaxation time $T_1 = 100\,\mu\mathrm{s}$ (with $T_1^{(|2\rangle)} \approx 50\,\mu\mathrm{s}$), pure dephasing $T_2^* = 30\,\mu\mathrm{s}$, and anharmonicity $\alpha/2\pi = -0.35\,\mathrm{GHz}$. 
As summarized in Table~\ref{tab:device_params}, the operating frequencies correspond to the bare frequencies of the system. 
The static bare capacitive couplings are fixed by the hardware design: the qubit-coupler couplings are $g_{1c_1}/2\pi = g_{2c_1}/2\pi = g_{2c_2}/2\pi = g_{3c_2}/2\pi = 0.17\,\mathrm{GHz}$, and the adjacent qubit direct couplings are $g_{12}/2\pi = g_{23}/2\pi = 0.01\,\mathrm{GHz}$. 
The time-dependent effective couplings $\tilde{g}_{ij}$ are dynamically generated from these static bare parameters. 
Furthermore, while the direct next-nearest-neighbor coupling ($g_{13}$) is assumed negligible in the main text, realistic setups may exhibit finite stray capacitance. To verify the robustness of our protocol, we simulated the gate performance under finite $g_{13}$ values; the results confirm that after minor parameter re-optimization, the average fidelity robustly remains above $98\%$ (see Fig.~\ref{FigA2}).

For dynamic control, we employ flat-top flux pulses with $\tau_{\mathrm{ramp}} = 2\,\mathrm{ns}$ cosine-smoothed edges to minimize high-level leakage. Recognizing the finite bandwidth limitations of arbitrary waveform generators in real experiments~\cite{RevModPhys.93.025005}, we systematically evaluated the gate performance using longer ramp times of 5 ns, 10 ns, and 20 ns. Analogous to the $g_{13}$ analysis, we finely re-optimized the control parameters for each $\tau_{\mathrm{ramp}}$ configuration. During this re-optimization, the total gate duration was not strictly fixed but was allowed to slightly adjust to accommodate the extended edges. The overall trend showed a minor increase in the total gate time; however, we emphasize that even for the 20 ns ramp, the total duration never exceeded 173 ns. The results demonstrate that the average gate fidelity robustly exceeds 99\% across all tested cases, confirming that the protocol maintains its significant speed and fidelity advantages even under conservative bandwidth constraints. Specifically, the fidelity reaches over 99.4\% for both 5 ns and 10 ns edges, and remains above 99.05\% even with extended 20 ns edges. This indicates that our protocol is not fundamentally bottlenecked by strict instrumental bandwidth limits; rather, it remains highly compatible with standard experimental control capabilities while maintaining state-of-the-art performance.

\paragraph{Calibration of effective couplings.}
The relation between coupler frequency $\omega_c$ and the effective coupling $\tilde{g}_{ij}$ is calibrated via vacuum‑Rabi oscillations. To measure $\tilde{g}_{12}$, we prepare $|100\rangle$ with $Q_3$ and Coupler~2 detuned, then apply a flux pulse to Coupler~1. The resulting chevron pattern (Fig.~\ref{FigA1}a) yields the oscillation frequency $\Omega$ at resonance ($\Delta_{12}=0$). This oscillation corresponds to a vacuum Rabi process in the single-excitation subspace, where $|\tilde{g}_{ij}| = \Omega/2$. Thus we obtain $\tilde{g}_{12}$. Scanning the flux amplitude maps $\tilde{g}_{12}$ versus $\omega_{c_1}$ (Fig.~\ref{FigA1}b), which sets the coupler biases for target couplings of ${\approx}16.72\,\mathrm{MHz}$ (Stage~1) and ${\approx}15.82\,\mathrm{MHz}$ (Stage~2a). The same procedure for the $Q_2$–$Q_3$ pair calibrates $\tilde{g}_{23}$ to ${\approx}14.85\,\mathrm{MHz}$ (Stage~1) and ${\approx}14.67\,\mathrm{MHz}$ (Stage~2b). The corrective CPhase gates in Stages~2a and~2b operate on the respective $|11\rangle \leftrightarrow |02\rangle$ resonance at these calibrated couplings.

\paragraph{Phase correction via Ramsey interferometry.}
Residual static $ZZ$ and $ZZZ$ interactions during Stage~1 introduce parasitic dynamical phases. We cancel these within the active cancellation stage (Stage 2) using two sequential CPhase gates (Stages 2a and 2b), with their phase shifts determined by a differential Ramsey protocol.

To extract a two‑body phase $\varphi_{ij}$, we compare Ramsey fringes on the target qubit with the control qubit in $|0\rangle$ (baseline) and $|1\rangle$ (probe). For $\varphi_{12}$, Ramsey measurements on $Q_2$ with $Q_1$ in $|0\rangle$ and $|1\rangle$ give
\begin{equation}
	\varphi_{12} = \varphi_{|1+0\rangle} - \varphi_{|0+0\rangle},
\end{equation}
where $\varphi_{|0+0\rangle}$ removes single‑qubit frame rotations. Analogous measurements give $\varphi_{23}$ and $\varphi_{13}$. The cancellation pulses in Stages 2a and 2b are then tuned to satisfy $\varphi^{\mathrm{corr}}_{ij} = -\varphi_{ij}$.

\paragraph{Fine‑tuning the three‑body interaction.}
The operating point of Stage~1 is fine‑tuned to minimize residual population in $|011\rangle$ and $|110\rangle$. While the Schrieffer–Wolff condition $\tilde{\omega}_1 + \tilde{\omega}_3 \approx 2\tilde{\omega}_2 + \alpha_2$ provides a starting point, finite pulse bandwidth and AC‑Stark shifts require empirical adjustment of the $Q_2$ detuning $\Delta$ and the coupler‑mediated interaction strength. This tuning reduces the infidelity contribution from residual non‑adiabatic population to $2.08 \times 10^{-3}$ (Sec.~\ref{Sec4}). The resulting optimal frequencies define the Stage~1 parameters in Table~\ref{tab:device_params}.

\section{Error Budget and Mechanistic Analysis}
\label{AppC}

This appendix details the physical origins of the error budget summarized in Table~\ref{tab:error_budget}. The infidelity is categorized into coherent control errors, incoherent decoherence, and leakage outside the computational subspace.

\paragraph{Coherent control errors.}
The total coherent error is $3.28 \times 10^{-3}$, arising from imperfections in the unitary evolution.
\begin{itemize}
	\item \textit{Residual non‑adiabatic population} ($3.08 \times 10^{-3}$): This dominant error comes from population left in the intermediate states $|011\rangle$ and $|110\rangle$ after the gate. These states are part of the computational (qutrit) basis and serve as virtual mediators for the three‑body coupling $J$. In a fast gate ($165\,\mathrm{ns}$), the adiabatic condition is relaxed, and their finite detuning from the $|101\rangle$–$|020\rangle$ manifold leads to non‑negligible diabatic mixing. The residual population is a direct measure of this non‑adiabaticity.
	\item \textit{Phase miscalibration} ($0.20 \times 10^{-3}$): A small residual error from imperfect cancellation of dynamical phases, reflecting the practical limit of the active cancellation protocol.
\end{itemize}

\paragraph{Incoherent errors.}
Incoherent errors account for $3.09 \times 10^{-3}$ of the infidelity, stemming from energy relaxation ($T_1$) and pure dephasing ($T_2^*$). A distinctive aspect of our protocol is the transient occupation of the second excited state $|020\rangle$. We model the transmon relaxation with harmonic scaling, $\Gamma_{|2\rangle\to|1\rangle} \approx 2\Gamma_{|1\rangle\to|0\rangle}$, which gives $T_1^{(|2\rangle)} \approx 50\,\mu\mathrm{s}$ given the base $T_1 = 100\,\mu\mathrm{s}$. This accelerated decay of the $|2\rangle$ state sets a fundamental floor for operations that involve higher levels, even with a short gate duration.

\paragraph{Leakage to higher levels.}
Leakage error contributes $2.61 \times 10^{-3}$. We define leakage strictly as population transfer to energy levels outside the modelled qutrit subspace ($n \ge 3$), e.g., $|030\rangle$, distinct from the intentional excursion to $|020\rangle$. Simulations that include up to 10 levels per transmon confirm this as the third‑largest error channel. The transmon anharmonicity provides some protection, but the finite bandwidth of the fast control pulses induces off‑resonant driving to these higher levels.

\paragraph{Summary.}
The total infidelity ($8.98 \times 10^{-3}$) is distributed almost equally among coherent control ($3.28 \times 10^{-3}$), incoherent decoherence ($3.09 \times 10^{-3}$), and leakage ($2.61 \times 10^{-3}$). This balance indicates that the gate performance is concurrently limited by control imperfections, finite coherence times, and leakage induced by the pulse bandwidth. The analysis confirms that the dominant coherent error (non‑adiabatic population) is intrinsic to the speed of the geometric phase accumulation, while the incoherent error is dominated by the enhanced relaxation of the $|2\rangle$ state. Further improvements would therefore benefit from advanced pulse shaping to reduce both non‑adiabatic transitions and leakage, alongside materials and design advances that extend $T_1$ and $T_2^*$.
\bibliography{paper.bib}  
\end{document}